\documentclass{article}
\usepackage{emulateapj,epsf}

\tighten
\newcommand{\kms}{$\,{\rm km\,s^{\scriptscriptstyle -1}}$}
\newcommand{\gtsim}{\ {\raise-0.5ex\hbox{$\buildrel>\over\sim$}}\
}
\newcommand{\ltsim}{\ {\raise-0.5ex\hbox{$\buildrel<\over\sim$}}\
}

\def\simlt{\lower.5ex\hbox{$\; \buildrel < \over \sim \;$}}
\def\simgt{\lower.5ex\hbox{$\; \buildrel > \over \sim \;$}}
\def\Sec{${}^{\prime\prime}$\llap{.}}

\slugcomment{Ap. J., accepted}
\lefthead{NEWMAN ET AL.}
\righthead{Cepheid Distance to NGC~4603}

\begin{document}

\title{A Cepheid Distance to NGC~4603 in Centaurus}

\author{Jeffrey A. Newman\altaffilmark{1}, Stephen
E. Zepf\altaffilmark{2}, Marc Davis\altaffilmark{1}, Wendy
L. Freedman\altaffilmark{3}, Barry F. Madore\altaffilmark{4}, \\
Peter B. Stetson \altaffilmark{5}, N. Silbermann\altaffilmark{6}, and Randy
Phelps\altaffilmark{3}}

e-mail: {\tt jnewman@astro.berkeley.edu, zepf@astro.yale.edu,
marc@astro.berkeley.edu, wendy@ociw.edu, \\
barry@ipac.caltech.edu,Peter.Stetson@hia.nrc.ca, nancys@ipac.caltech.edu,
rphelps@physics.oberlin.edu}

\vskip 24pt

\altaffiltext{1}{Department of Astronomy, University of
California, Berkeley, CA 94720}
\altaffiltext{2}{Department of Astronomy, P.O. Box 208101, Yale University, New
Haven, CT 06520}
\altaffiltext{3}{Observatories of the Carnegie Institute of Washington, 813 Santa Barbara St.,Pasadena, CA 91101}
\altaffiltext{4}{NASA/IPAC Extragalactic Database, Infrared Processing and Analysis Center, Jet Propulsion Laboratory, California Institute of Technology, MS 100-22, Pasadena, CA 91125}
\altaffiltext{5}{Dominion Astrophysical Observatory, 5071 W. Saanich Rd., Victoria, B.C., Canada V8X 4M6}
\altaffiltext{6}{Jet Propulsion Laboratory, California Institute of Technology, MS 100-22, Pasadena, CA 91125}

\begin{abstract}

In an attempt to use Cepheid variables to determine the distance to
the Centaurus cluster, we have obtained images of NGC~4603 with the
Hubble Space Telescope for 9 epochs (totalling 24 orbits) over 14
months in the F555W filter and 2 epochs (totalling 6 orbits) in the
F814W filter.  This galaxy has been suggested to lie within the
``Cen30'' portion of the Centaurus cluster, which is concentrated
around a heliocentric redshift of $\approx 3000$ \kms, and is the most
distant object for which this method has been attempted.  Previous
distance estimates for Cen30 have varied significantly and some have
presented disagreements with the peculiar velocity predicted on the
basis of full-sky redshift surveys of galaxies, motivating our
investigation.  Using our WFPC2 observations, we have found 61
candidate Cepheid variable stars with well-determined oscillation
periods and mean magnitudes; however, a significant fraction
of these candidates are likely to be nonvariable stars whose magnitude
measurement errors happen to fit a Cepheid light curve of significant
amplitude for some choice of period and phase.  Through a maximum
likelihood technique, we determine that we have observed $43 \pm 7$
real Cepheids (with zero excluded at $>9 \sigma$) and that NGC~4603
has a distance modulus of $32.61_{-0.10}^{+0.11}$ (random, 1 $\sigma$)
$^{+0.24}_{-0.25}$ (systematic, adding in quadrature), corresponding
to a distance of $33.3^{+1.7}_{-1.5}$ (random, 1 $\sigma$)
$^{+3.8}_{-3.7}$ (systematic) Mpc.  This result is consistent with a
number of recent estimates of the distance to NGC~4603 or Cen30 and
implies a small peculiar velocity consistent with predictions from the
$IRAS$ 1.2 Jy redshift survey if the galaxy lies in the foreground of
the cluster.

\end{abstract}



\section{Introduction}

The gravitational field of the inhomogeneous distribution of mass in
the Universe produces observable deviations from the smooth Hubble
expansion.  Well-determined distances to galaxies provide an
opportunity to measure their motions relative to the ``Hubble flow''
-- so-called peculiar velocities -- which can lead to mass estimates
for a variety of systems, as in studies of the Local Group and of the
Virgocentric infall, or on larger scales via comparisons of peculiar
velocity measurements to expectations from full-sky redshift surveys (Dekel
1994, Willick \& Strauss 1995).

In such analyses, the Centaurus region is probably the most perplexing
zone of large-scale flow in our vicinity.  It has a complex spatial
structure, and its peculiar velocity has been measured in some studies
to be much higher than that expected from the observed galaxy density.
Lucey, Currie, and Dickens (1986b) first called attention to the
apparently bimodal nature of the Centaurus cluster at ({\it l,b}) =
(302$^\circ$, 22$^\circ$), dividing it into two pieces at apparent
redshifts in the Local Group (LG) reference frame of approximately 2800 and
4300 \kms\ (Cen30 and Cen45, respectively).  In a deeper study of the
central portions of the cluster, Stein et al. (1997) found that dwarf
galaxies in Centaurus exhibit a clear concentration around the
redshift of NGC~4696 (an elliptical galaxy which is the brightest in
the cluster), $v_{lg}=2674 \pm 26$ \kms, tracing a galaxy cluster
of velocity dispersion $933 \pm 118$ \kms\ that they identify with
Cen30.  Cen45, they determined, more strongly resembles a group falling
into Cen30, with a small velocity dispersion ($131 \pm 43$ \kms) and a
population dominated by late-type galaxies.  

A number of secondary distance
indicators have by now been applied to Centaurus galaxies, with often
contradictory results.  Aaronson {\it et al.\/} (1989) were the first
to obtain distances for Centaurus spiral galaxies; they measured
peculiar velocities of $-80\pm 250$ and $+10\pm 450$ in the Local
Group reference frame for Cen30 and Cen45, respectively.
However, obtaining Tully-Fisher distances to these clusters is somewhat
problematic.  Because of the large velocity dispersion of Cen30 and
the relatively small number of galaxies in Cen45, separating cluster
from background spirals is very difficult (see Lucey, Currie and
Dickens 1986a and Giovanelli et al. 1997 for examples).  Reflecting
these difficulties, Aaronson et al. identify 6 galaxies spanning
nearly 2 magnitudes in distance modulus as belonging to Cen45.  In contrast, based on $D_n - \sigma$ observations of elliptical galaxies, Faber et
al. (1989) reported the peculiar velocities for Cen30 and Cen45 to be
$+527\pm 214$ \kms\ and $+1090\pm 336$ \kms\ in the Local Group frame.

In an attempt to resolve such contradictory estimates of the distance
to the Centaurus region, we have undertaken a search for Cepheids in
the spiral galaxy NGC~4603 to firmly establish its location.  This
galaxy is located near the center of the Cen30 cluster in position on
the sky, and has a velocity $v_{lg}=2321 \pm 20$ \kms\ (Willick et
al.), well within the velocity dispersion of the cluster.  NGC~4603
has an inclination of 53$^\circ$ and a 21cm width (20\%) of 411 \kms\
with isophotal ($D_{25}$) diameter of 1.6'; Aaronson {\it et al.\/}
(1989) show that it fits onto their IRTF relation for Cen30 galaxies
quite well, with a distance modulus within 0.07 magnitudes
(0.3$\sigma$) of that derived for the cluster.  It thus seems an
appropriate choice for such a study.  Such a study should also allow
tests of the validity of the uniformity of the Tully-Fisher or $D_n -
\sigma$ relationships to a greater distance than has been possible
before.

However, even the smaller estimates of the distance to Cen30 place it
substantially further than any galaxy for which a search for Cepheids
has been previouly attempted, even using the Hubble Space Telescope;
the greatest distance modulus previously measured with this method is
that to NGC~4639, $32.03\pm 0.22$ ($25.5\pm2.5$ Mpc; Saha et
al. 1997).  The redshift of Cen30 is roughly twice that of the Virgo
or Fornax clusters.  It is thus reasonable to expect that observing
Cepheids in NGC~4603 should be difficult; not only do more distant
Cepheids appear fainter, but also the crowding of stars that
complicates photometry becomes more severe as the angular size
distance increases.  Furthermore, the Centaurus cluster lies behind a
zone of substantial ($A_V\sim 0.5$) Galactic extinction, making any
stars observed that much fainter.  In this regime, photometric errors
are significant enough that nonvariable stars have an appreciable
probability of appearing to vary in a manner indistinguishable from
a Cepheid with significant amplitude.  Such obstacles might be overcome
by observing at many more epochs or with a greater exposure time per
epoch than in prior Cepheid studies, but the limited availability of
HST makes that infeasible.

Therefore, we have developed new techniques for dealing with such a
dataset.  Instead of relying on a set of variability criteria for
preselection, we attempt to fit template Cepheid light curves to all
stars with well-determined photometry and then apply a series of
criteria that are effective at eliminating nonvariables.  Even that
technique leaves a substantially contaminated list of candidate
Cepheids.  We therefore do not obtain distance moduli from a direct
Period-Luminosity relation fit, but rather have developed a Maximum
Likelihood formulation that accounts for the properties of
nonvariables that mimic Cepheids and of the probability of selecting
an actual Cepheid of given properties based upon the results of
realistic simulations.  These techniques allow us to minimize the
biases in distance determination that might otherwise appear and which
may have affected other Cepheid studies that have
pushed the limits of the technique.  

We describe the details of the observations in $\S$ 2, the
procedures used to analyze the data and find Cepheids and our
simulations thereof in $\S$ 3, and our Maximum Likelihood formalism
and the determination of the distance to NGC~4603 in $\S$ 4.

\section{Observations}

We have observed NGC~4603 using the Wide Field and Planetary
Camera 2 (WFPC2) instrument on the Hubble Space Telescope (HST).
HST made a total of 11 distinct visits to the targeted field: 9,
totaling 24 orbits, using the F555W filter (roughly equivalent to
Johnson $V$), and 2, totaling 6 orbits, using the F814W filter
(similar to Kron-Cousins $I$).  To ensure ease of data analysis, the
same orientation was maintained for all observations; the
telescope was generally dithered by 5.5 planetary camera pixels
($\approx$ 0\Sec25) between orbits.  Two successive frames of
data were obtained during each orbit to minimize the effects of
cosmic rays.  Due to technical limitations (such as the time
required to acquire the target field and the limited visibility of
NGC~4603 during the course of an orbit), the total integration
time was 900-1300 seconds per frame.

Our observing strategy was in general similar to that used for the
$H_0$ Key project (see, e.g., Freedman et al. 1994); however, due
to the large predicted distance of NGC~4603 ($>$20 Mpc), we could
expect to find only the longest period Cepheids (i.e., $P \simgt
25$ days). In fact, if NGC~4603 were located at $\simgt$45 Mpc,
Cepheids in this galaxy would be too faint to discover at all with
the WFPC2 instrument.  We thus tried to optimize our observing
sequence to facilitate the discovery of longer-period variables.
Our original plan was to perform 8 F555W visits over the course of
$\approx 60$ days in 1996, spaced to maximize our ability to
detect and parameterize Cepheids with a variety of periods (as
described in Freedman et al. 1994).  Unfortunately, the final
observation planned for 1996 did not occur due to an HST safing
event.  Our sensitivity for the longest-period Cepheids --- exactly
those which are brightest and easiest to find --- is therefore
limited; those detected suffer from substantial aliasing in period
determination.  Details of the observations performed are listed
in Table 1.  In Figure 1, we show the results of a simulation for
the expected error in period determination as a function of period
for the sampling ultimately used, illustrating the effects of
aliasing.

\section{Data Analysis}

\subsection{Photometry}

The data were calibrated via the standard Space Telescope Science
Institute pipeline processing (Holtzman et al. 1995), applying the
Hill et al. (1998) long-exposure magnitude zero point.  Each frame
was also corrected for vignetting and geometrical effects on the
effective pixel area as described in Stetson et al. (1998).\\

\vbox{%
\begin{center}
\leavevmode
\hbox{%
\epsfxsize=8.9cm
\epsffile{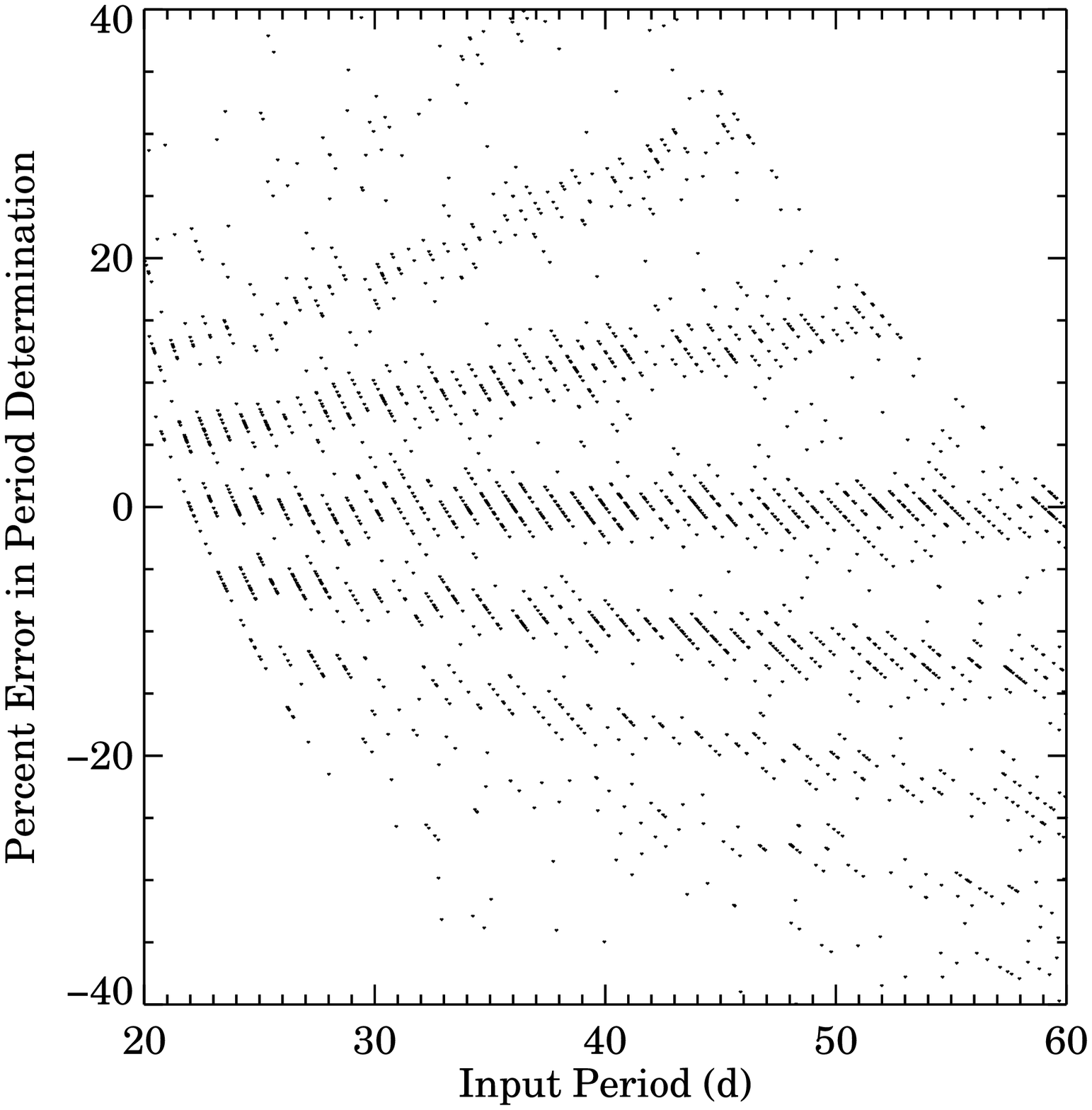}}
\begin{small}
\figcaption{\small The percentage difference between the period measured using
our algorithms and the actual period for simulated observations of
stars with Cepheid light curves of a variety of phases and amplitudes
measured (with realistic magnitude errors appropriate for stars with
$F555W$=27.3-27.5) at the epochs of our actual observations, as a
function of period.  The period errors are dominated by mistakenly
identifying an alias as the actual period.  The diagonal striping
apparent here is due to the gridding in period used for fitting
template light curves.}
\end{small}
\end{center}}

Background levels in the data frames were high enough and exposure
times long enough that neither charge transfer inefficiencies nor
variations in the photometric zero points with exposure time
should significantly affect our results (cf.  Rawson et al. 1997
and references therein).  Each of the
WFPC2 chips was analyzed separately.  Because the second WF chip
contained the nucleus of NGC~4603, crowding was severe and few
stars could be resolved in it; that chip was therefore omitted
from analysis.  The fourth WF chip was directed at an outer
portion of the galaxy, containing few stars and no significant
numbers of Cepheids; it, too, was therefore removed from our
analysis.

Photometry was then performed on each of the data frames using the
DAOPHOT II/ ALLFRAME package (Stetson 1987).  As an independent check,
magnitudes were also obtained using a version of DoPHOT (Schechter,
Mateo \& Saha 1993) modified by Abi Saha for use with HST data (see,
e.g., Ferrarese et al. 1996).  The DoPHOT reductions were used as a
consistency check only; the analysis presented in this paper is based
on the ALLFRAME photometry alone.  For F555W observations, the two
sets of photometry agreed to within $\pm$ 0.08 magnitudes on average;
this agreement is consistent with that found for distant galaxies
observed as part of the Key Project (e.g., Ferrarese et al. 1996,
Silbermann et al.  1998).  The ALLFRAME analysis was more extensive
and resulted in larger numbers of Cepheid candidates; for candidates
found using both packages, the agreement in period was found to be
well within the errors quoted below.

For the ALLFRAME photometry, procedures similar to those of the HST
Key Project on the Extragalactic Distance scale were used (see, e.g.,
Kelson et al. 1996 for a more detailed description).  ALLFRAME
performs photometry by fitting a predefined point-spread function
(PSF) to all stars on a frame and iteratively determining their
magnitudes.  Files describing the WFPC2 PSF and its variation across
the field (determined from observations of globular clusters; cf. Hill
et al. 1998) were provided by P.  Stetson.  For each epoch, up to six
HST frames were obtained, and thus up to six measurements of each
star's magnitude were made.  Those measurements are sometimes
contaminated by cosmic rays or other transient phenomena.  Although
ALLFRAME attempts to limit their effect, it was found that simply
averaging the magnitudes determined using ALLFRAME and weighting them
according to their error estimates sometimes yields very inaccurate
results.  We therefore experimented with a number of robust estimators
for the mean of the ALLFRAME measurements (including median, Tukey
biweight, and trimean; cf. Beers et al. 1990) using the magnitudes of
artificial stars inserted (using the ALLFRAME PSF) on our data
frames. The most successful proved to be an iterative reweighting
method described by Stetson (1997).  In this technique, each
measurement's weight is altered according to its difference from the
prior estimate of the mean (taken to be the median of the epoch's
measurements for an initial guess), as implemented here according to
the formula:

\begin{equation}
\sigma_i'^2= {{\sigma_i^2} \over{1+({{m_i-\bar{m}} \over{2
\sigma_i}})^2}},
\end{equation}

\noindent where $m_i$ is the $i$th measurement, $\sigma_i$ is the error
estimate in that quantity after the prior iteration, and $\bar{m}$
is the estimate of the mean from the prior iteration; after this
adjustment of the weights, a new determination of the mean is
made.

An estimate of the uncertainty in each epoch's mean magnitude
measurement was obtained from the weighted standard deviation of
the data:
\begin{equation}
\sigma_m^2 = {{\sum{{(m_i-\bar{m})^2} \over{\sigma_i^2}} \over{{\sum{{1} \over{\sigma_i^2}}} (n-1)}}},
\end{equation}
where $n$ is the total number of measurements used in determining
$\bar{m}$.  The resulting uncertainty estimates were generally
accurate to 10-20
\% (based upon the median $\chi^2$ of the comparison of each epoch's
magnitude measurements for a star to the mean magnitude obtained from
all F555W measurements for that star).

An additional potential source of photometric errors is the estimation
of the background counts underlying the star (due to unresolved stars,
H II regions, etc.).  ALLFRAME estimates that background level by
taking the median number of counts from pixels within some annulus
about the star whose magnitude is being measured.  Initially, our
studies were done using an annulus from 3 to 20 pixels in radius from
the stars; we later performed photometry using background annuli from
3 to 10 pixels and from 3 to 6 pixels.  For $F555W$ observations of
faint stars with well-determined photometry, the mean change in
epochal magnitudes was 0.000 $\pm 0.001$ mag, the RMS 0.10 mag, and
the root median square difference (also known as the probable error) 0.047 mag when photometry was done
with a 3-10 pixel radius sky annulus instead of 3-6.  For $F814W$, the
corresponding numbers were 0.000 $\pm$ 0.003 mag, 0.18 mag, and 0.079
mag.  Somewhat larger differences resulted from changing from a 3-10
pixel background annulus to 3-20, though increasing the background
region does reduce the scatter among the magnitude measurements for a
given star. Therefore, for the mean magnitudes presented here, we have
adopted the 3-10 pixel background level and included the probable magnitude error within the uncertainty estimate for each star's magnitude.

We found that using the ALLFRAME error estimates for weighting when
averaging magnitudes for a given star yielded a systematic bias
towards the brighter measurements.  This bias is minimal when errors
are small, but is several tenths of a magnitude for the faintest
stars.  We have chosen to perform averaging of magnitudes rather than
fluxes as it yielded a lower scatter of epochally averaged magnitudes
in tests of both artificial and actual stars, and our ability to
discriminate variations in brightness from the effects of magnitude
measurement errors was a major limiting factor in this work; a much
smaller but significant bias in the opposite sense was also found for
flux averaging.  A comparison of the averaged $F555W$ magnitudes for
stars on Chip 1 to their unbiased median magnitude measurements may be
found in Figure 2, along with a functional fit to the bias (here and
in the remainder of the paper, $F555W$ and $F814W$ will refer to magnitudes
obtained by combining ALLFRAME measurements with an appropriate zero
point; no aperture or bias corrections have been applied to them.  $V$
and $I$ will be used to refer to fully corrected magnitudes on the
Johnson and Kron-Cousins systems, respectively).  Such functional fits
were used in a Brent's method-based algorithm (cf. Press et al. 1992)
to remove the biases in mean $V$ and $I$ magnitudes before color
measurements or comparison to Cepheid P-L relations.  Any biases due
to averaging procedures should be corrected via this method. The
expected error in the amount of the bias correction due to errors
in measuring a star's mean magnitude is much less than the width of
the P-L relation in both $V$ and $I$ ($\sim$ 0.04 mag for typical
candidate Cepheids in our dataset), and thus should have no effect on
our final results.

\subsection{Cepheid Identification}

NGC~4603 is the most distant galaxy for which a Cepheid search has
been attempted; the required photometry presented a considerable
challenge.  Because the errors in the magnitude measurements for each
epoch were a substantial fraction of typical Cepheid amplitudes and
because of the limited number of epochs available, common techniques
for identifying variables (see, e.g., Rawson et al.  1997 and
references therein) proved to be of limited utility; for instance, the
phase dispersion minimization method, which requires binning the
observations in phase, is hardly optimal for noisy datasets with such
a limited number of observations (Stellingwerf 1978).  We instead have
adopted an alternative approach loosely based on that described in
Stetson 1996.

\vbox{%
\begin{center}
\leavevmode
\hbox{%
\epsfxsize=8.9cm
\epsffile{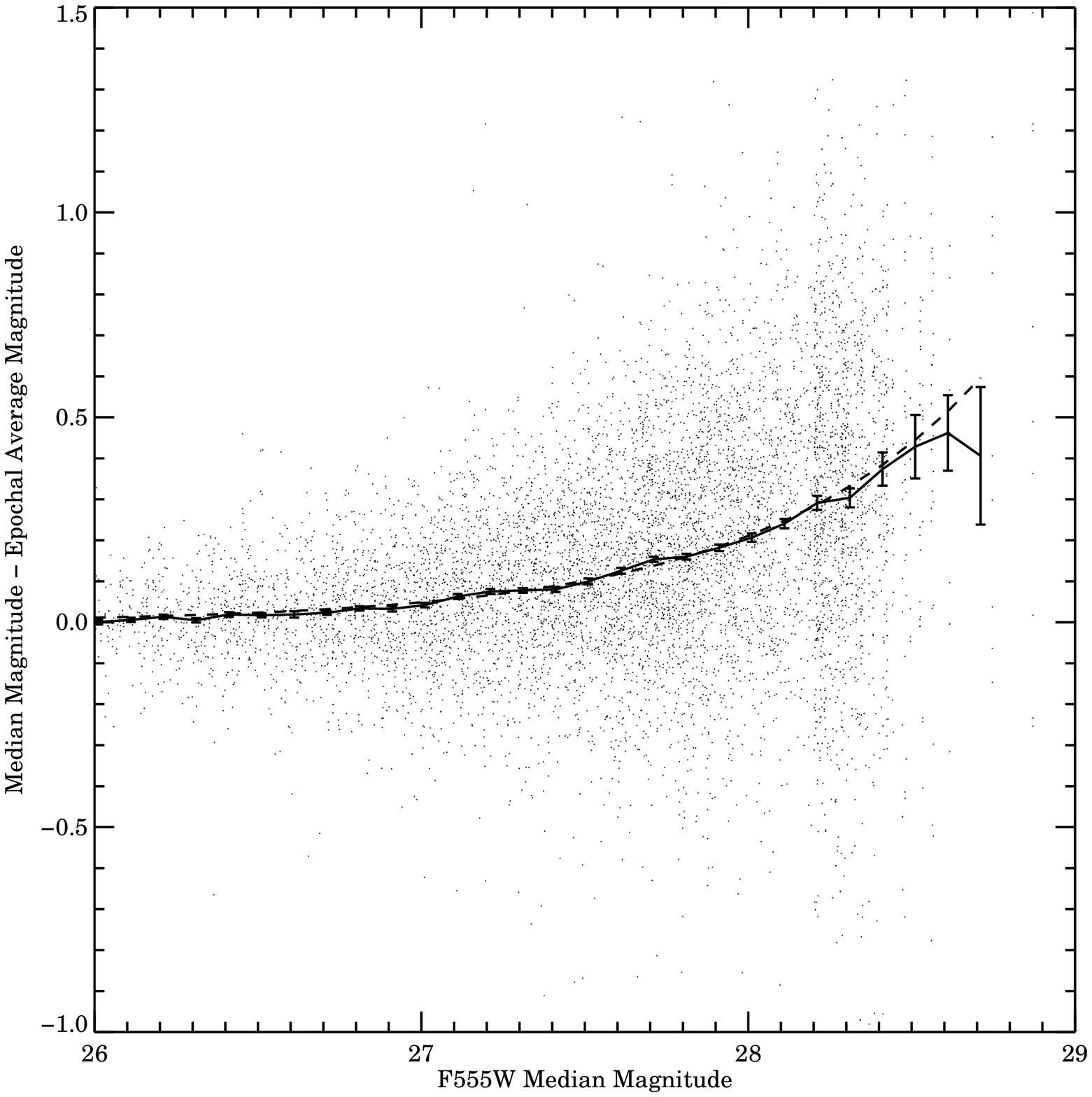}}
\begin{small}
\figcaption{\small A plot of the difference between the unbiased
median magnitude measurement for a star and the biased epochal average
measurements.  It is readily apparent that this bias is greater at
fainter magnitudes.  The solid line traces the median bias for stars
in 0.1 mag wide bins, with error bars corresponding to the standard
error of the mean for each bin; the dashed line is a regression fit to
that data, showing that the bias is well represented by a power law in
the actual (as opposed to measured) flux.  So long as there exists a
one-to-one correspondence between the unbiased actual and biased
measured magnitudes (which is true for measured $F555W < $ 28.125,
$F814W < $ 27.166), we may use this fit relation to correct for the
bias.}
\end{small}
\end{center}}

The computing power of modern workstations is now sufficient that we
could attempt to fit the photometry for every well-observed star to a
grid of model Cepheid light curves (taken from Stetson 1996) with a
wide range of periods and phases (in general, we sampled the period in
1 day increments and phase in increments of 0.025 for our variable
search). This reduces the problem to a set of linear regressions to
obtain mean magnitude and amplitude, a quite rapid procedure.  By
minimizing $\chi^2$ on this grid, we obtain an estimate of the most
appropriate Cepheid light-curve parameters for a given star.
Nonvariable stars emerge from this fitting process with low
amplitudes, typically substantially smaller than the amplitude error
estimates resulting from the procedure; they can be rejected on this
basis.  For variables, the width of the minimum in the variation of
$\chi^2$ with period allows us to estimate our uncertainty in that
parameter for a given star.  We also confirmed our light-curve fits by
performing a nonlinear $\chi^2$-minimization fit to the data for
suspected variables with our best-fitting parameters as initial
guesses.  This generally resulted in minimal changes in parameters,
indicating that our grid was sufficiently fine.

There are complications for longer-period variables ($> 40$d), for
which multiple deep minima in $\chi^2$ appear due to aliasing.
However, our Monte Carlo analysis (see $\S 3.3.2$) indicates that we
still determine the periods of such Cepheids to 10-20 \% accuracy
(with the period errors then dominated by the spacing between the
minima, reflecting the possibility that the deepest $\chi^2$
minimum occurs at an alias --- typically the nearest one --- of the
actual period, as reflected in Figure 1).

Stetson (1996) also defines model Cepheid $I$-band light curves
based upon the same parameters as those for $V$.  Thus, once
a $V$-band fit is obtained, two determinations of the mean $I$
magnitude for a star can be made by combining our two epochs'
magnitude measurements and the expected $I$ variation at the phase
of those measurements (a method not unlike that described in
Sandage et al. 1997).  We used a weighted mean of these two
determinations to estimate the mean $I$ magnitude for our
variables.

\subsection{Simulations of Cepheid Detection Rates and Expected
Errors}

Because the Cepheids we are looking for are so faint, it is critical
to confirm our ability to unambiguously detect such stars and to limit
contamination of our sample of Cepheids with nonvariable stars.  We
therefore performed our variable search on two sets of artificial
photometric data, one consisting of intrinsically nonvariable stars
and one of stars changing in brightness (before measurement errors)
according to template Cepheid light curves.  For each star in one of
these datasets, artificial magnitude measurements were made according
to the actual timing of the HST visits.  To account for the
possibility of non-Gaussian distributions of errors, these constant or
varying light curves were modified by numbers selected randomly from
the set of actual deviations of $F555W$ or $F814W$ magnitude
measurements of stars in a given magnitude range from their overall
robustly determined mean magnitude (which, having been found from 48
or 12 magnitude measurements, respectively, were much more accurate
than a single-frame measurement).  To retain the information on a
measurement's quality present in the ALLFRAME error estimates, each
frame's magnitude measurement in the artificial datasets was assigned
the magnitude uncertainty estimate from the appropriate star and frame
number for the measurement error applied.  This analysis was performed
for stars in ten 0.2 magnitude wide ranges, equivalent to $F555W$
magnitudes from 26.3-26.5 to 27.7-27.9.

\subsubsection{False Positives}

Attempting to find variables in our fake photometry of nonvariable
stars generates candidate ``variables'' that mimic real Cepheids,
hereafter referred to as ``false positives.''  The rate of these
misidentifications in the NGC~4603 dataset is such that any reasonable
list of candidate Cepheids we may produce will be contaminated with
nonvariable stars.  However, using our simulations, we have been able
to find a number of criteria that can help reject such candidates.
Some results of these simulations are plotted in Figures 3-5.

Foremost, the majority of the false positives possess low amplitudes
($<$0.6 magnitudes peak-to-peak in the principal Fourier component,
the form of amplitude measured by our template fitting technique), as
illustrated by Figure 4, so excluding low-amplitude variables
eliminates many of them.  We also exclude stars with low amplitudes
compared to their statistical error estimates from least-squares
fitting.  A further test that proved very useful was to require all
candidate Cepheids to have at least four data points more than 1.2
$\sigma$ away from their robustly determined overall mean magnitude;
nonvariable stars rarely possessed that many deviating points.  This is
effectively a test for a non-Gaussian distribution of magnitude
measurements (a characteristic of Cepheid light curves) that is
resistant to a small number of outliers.  Another helpful restriction
was ruling out very short period ($< 24 $d) candidates, as those were
far more likely to be false positives than real Cepheids (due to the
increasing ability to make a given light curve match given magnitude
variations with some choice of phase at shorter periods).  All results
discussed in this paper utilize variable-finding routines that perform
all of these tests.  The number of nonvariable stars which survive our
variability criteria is fairly low; as shown in Figure 3, on Chip 1
(the Planetary Camera, which has the deepest effective photometry) we
find that $\sim 0.5$ \% of all faint stars with $F555W \simeq 27$ (but
up to 6 \% by $F555W=27.6$) may be misclassified as Cepheids in our
analysis.

\vbox{%
\begin{center}
\leavevmode
\hbox{%
\epsfxsize=7.5cm
\epsffile{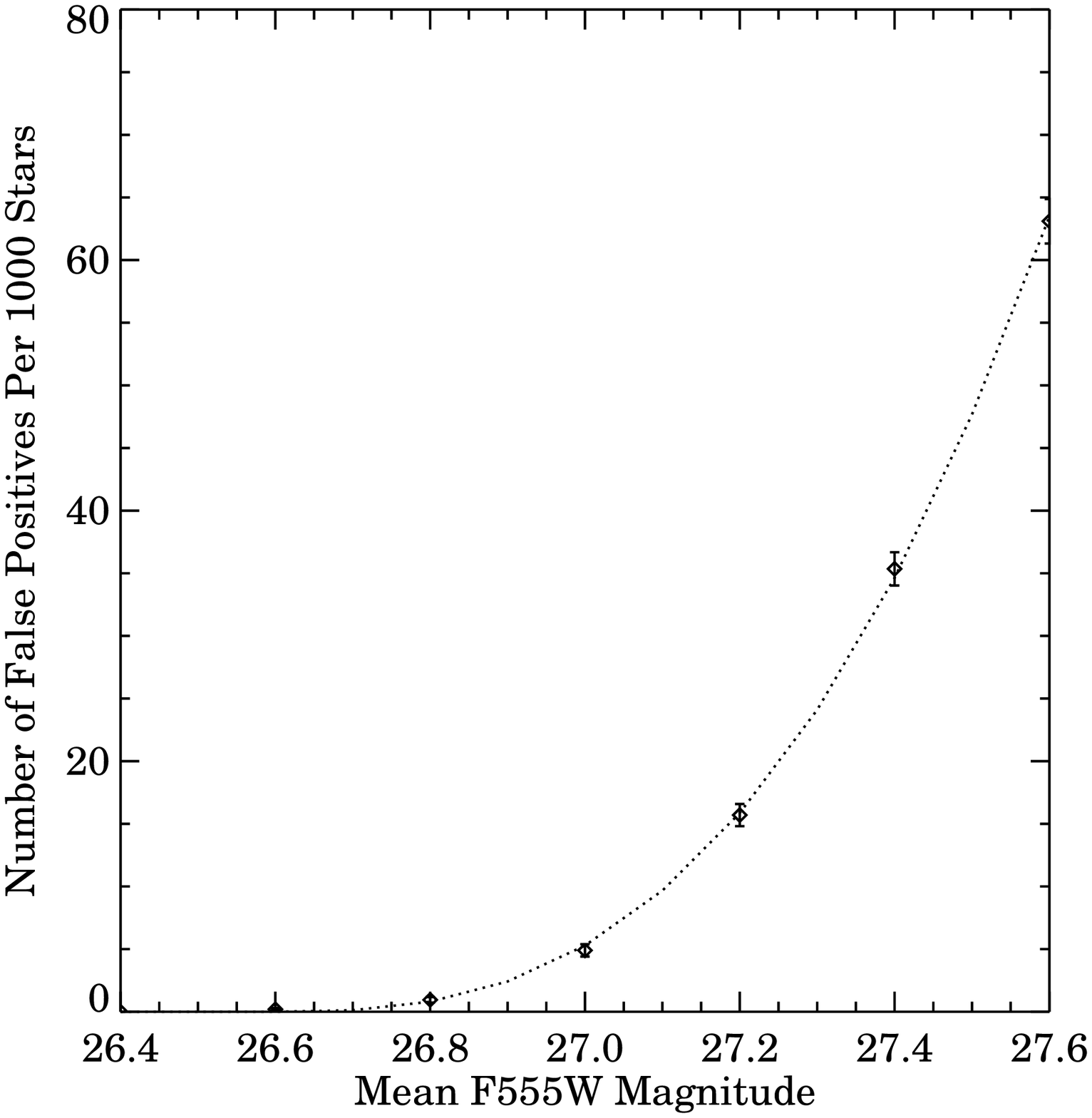}}
\begin{tiny}
\figcaption{\small Results of simulations for the rate at
which false positives occur for stars on Chip 1 (the Planetary
Camera) as a function of $F555W$ magnitude using our variability criteria, which have excluded the great majority of such misidentified stars.  In
this and all following figures, the dotted line indicates the fit
used in obtaining maximum likelihood estimates of distance (see $\S 3.1$).}
\end{tiny}
\end{center}}

Given the large numbers of faint stars in our dataset, it is likely
that our list of candidate Cepheids contains many which are actually
nonvariable.  There were roughly 3000 stars with well-determined
photometry (i.e., magnitude measurements on $>$ 90\% of all frames) on
Chip 3, which has the most stars found; there are roughly 2100 such
stars on Chip 1 (the Planetary Camera).  We searched for Cepheids
among these.  Integrating the false positive rate over our observed
magnitude distributions, we expect 34.0 $\pm ~5.8$ on Chip 1, and 56.9
$\pm ~7.5$ on Chip 3.  In contrast, 61 stars on Chip 1 (all of $F555W$
magnitude 27 or\\
\vbox{%
\begin{center}
\leavevmode
\hbox{%
\epsfxsize=7.5cm
\epsffile{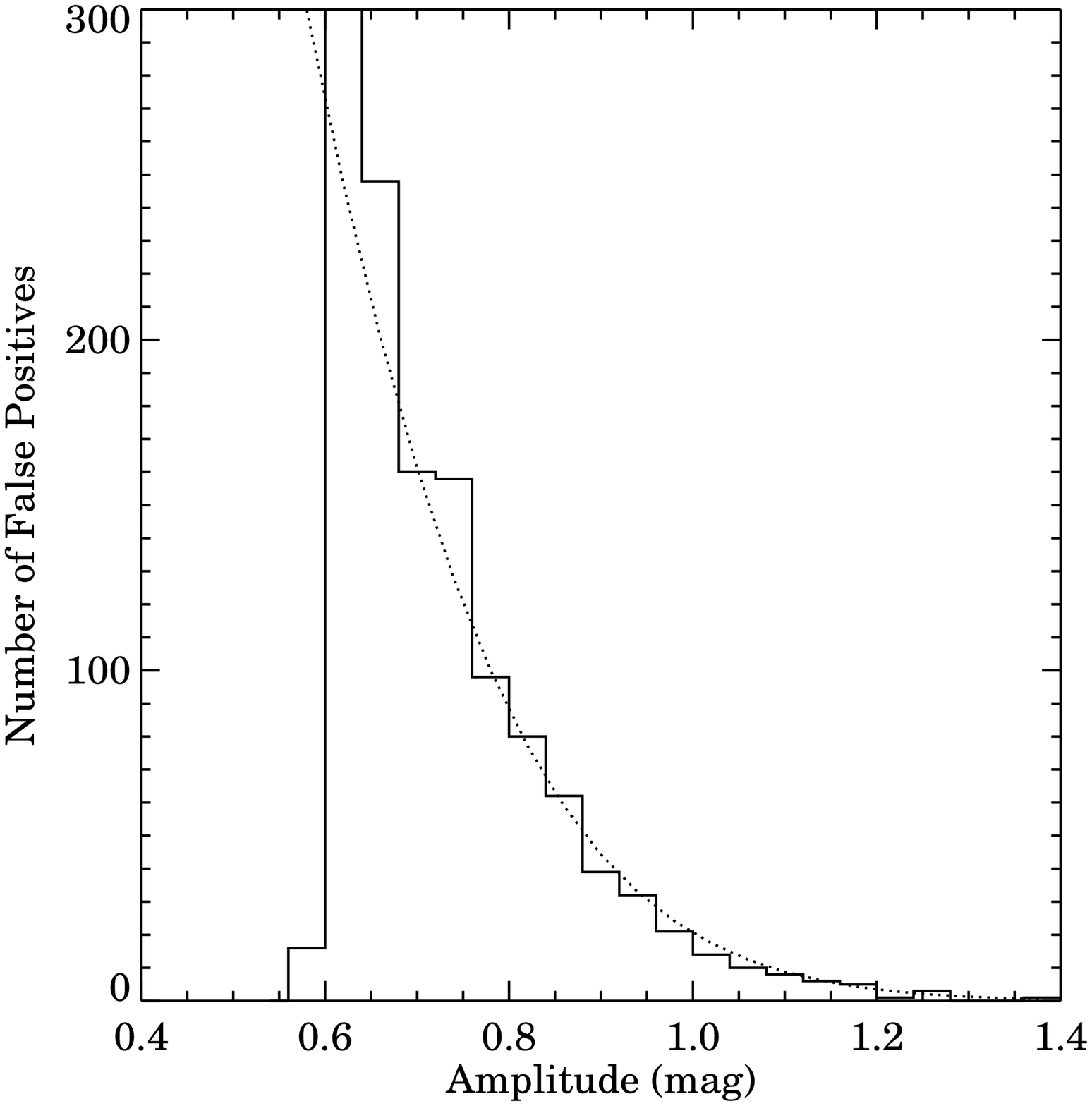}}
\begin{small}
\figcaption{\small Results of simulations for the distribution
of false positives in amplitude for magnitudes typical of our
Cepheid candidates.  Stars with measured amplitude below 0.60 mag
were rejected as Cepheid candidates.}
\end{small}
\end{center}}
 fainter) and 69 on Chip 3 passed all our Cepheid
detection tests.  Therefore, we concluded that the set of putative
variables on Chip 3 is too contaminated to yield useful information,
and we have concentrated on the candidate Cepheids on Chip 1 for further
analysis.

\subsubsection{Artificial Cepheids}

To determine our ability to detect any variable stars present in our
dataset, we generated data with realistic photometric errors determined as
described above applied to analytically defined Cepheid light curves
(Stetson 1996) with randomly selected periods, amplitudes, and phases.  These
``artificial Cepheid'' Monte Carlo simulations yielded encouraging
results.  On both the Wide Field and Planetary Camera chips, 45-70 \%
(depending upon input parameters; the recovery rate was substantially
less than this for candidates with input amplitudes below 0.6 mag, as
should be expected given our variability criteria) of those Cepheids
with mean $F555W \ltsim 27.5$ passed our tests, with probable
magnitude measurement errors of $\ltsim 0.1$ mag and period errors of
$\sim 10 \%$, quite comparable to the uncertainty estimates from our variable
search routines.  Some of the results of these simulations are
presented in Figures 6-9.

\section{Results}

A number of potential Cepheid variables on Chip 1 with well-
determined parameters were found.  Light curves for some of these candidates
are shown in Figure 10.  Their properties are summarized in Table 2.  Epochal photometry and light curves for all candidate Cepheids are available via WWW.
\footnote{\texttt http:$\slash\slash$www.astro.berkeley.edu/$\sim$marc/n4603/}

\vbox{%
\begin{center}
\leavevmode
\hbox{%
\epsfxsize=7.5cm
\epsffile{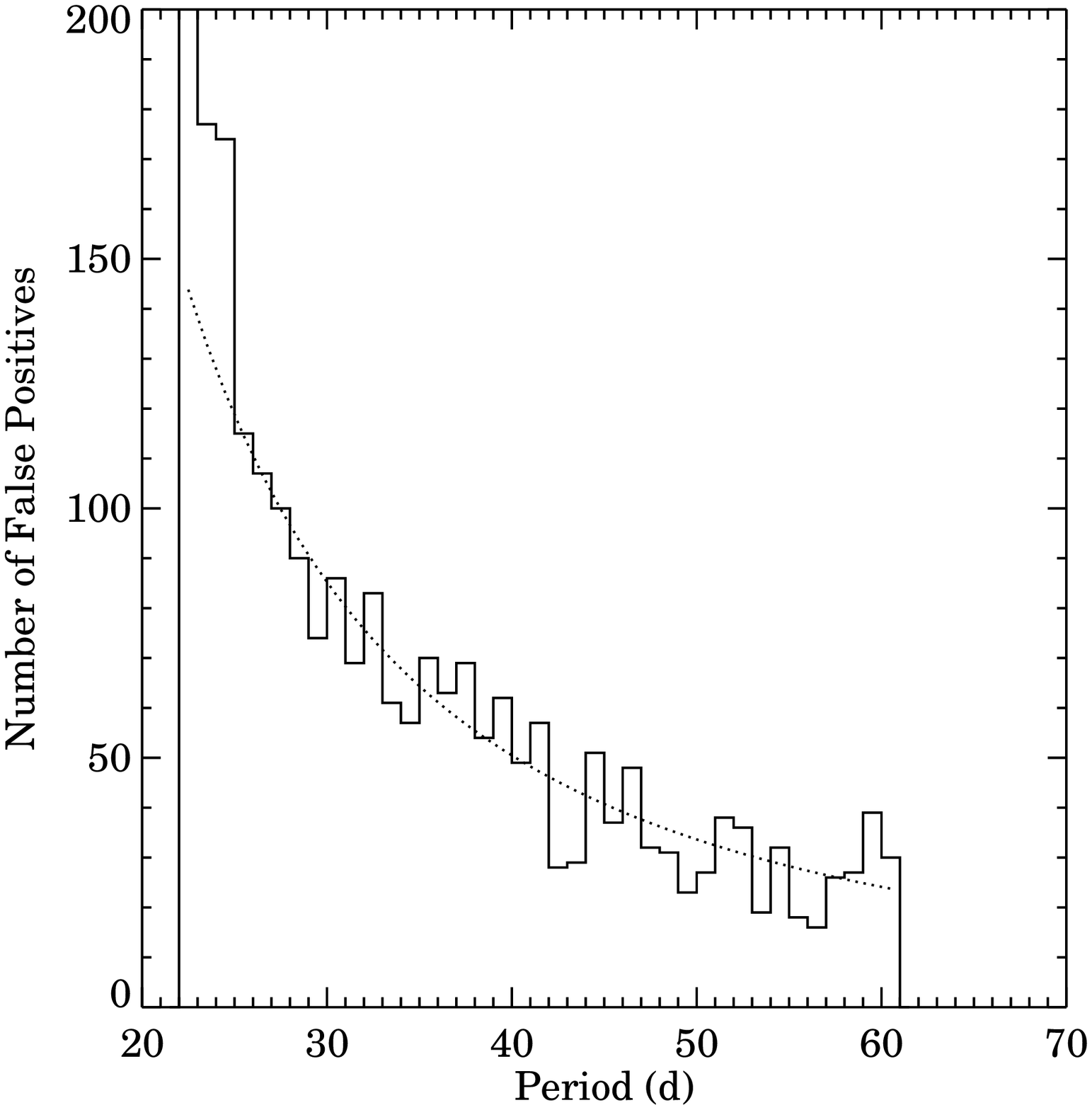}}
\begin{small}
\figcaption{\small Results of simulations for the distribution
of false positives in period, for magnitudes typical of our
Cepheid candidates.  Stars with measured period less than 24 d or
greater than 60 d were rejected as Cepheid candidates.}
\end{small}
\end{center}}

The $F555W$ and $F814W$ mean magnitudes of the variables determined
from the chi-squared minimization were converted to Johnson $V$ and
Kron-Cousins $I$ using equations from Hill et al. (1998):
\begin{eqnarray}
V = F555W-25&-&0.052(V-I) \\
\nonumber   &+&0.027(V-I)^2+22.510 \\
I=F814W-25&-&0.063(V-I) \\
\nonumber &+&0.025(V-I)^2+21.616, \\
\nonumber \end{eqnarray}
where $F555W$ and $F814W$ are the measured ALLFRAME magnitudes for
the corresponding filters.  Fixed aperture corrections of $-0.17
\pm 0.01$ magnitudes each, determined based on those obtained in
prior Key Project ALLFRAME analyses for the PC (Graham et al. 1998), were
also applied when obtaining the $V$ and $I$ magnitudes.

\subsection{Maximum Likelihood Analysis}

In order to extract as much of the information available from our set
of candidate Cepheids as we can despite the presence of false
positives, we have performed an extensive maximum likelihood analysis
to determine the distance modulus of NGC~4603.  This required
knowledge of our variable detection rates and the errors in measuring
the period and magnitude of actual Cepheids, in addition to the
distribution in period, magnitude, and amplitude of false positives;
these could all be obtained from our Monte Carlo simulations
(q.v. above).  To perform the maximum likelihood analysis, we also
required some knowledge of the distribution in period of actual
Cepheids; this was 
\vbox{%
\begin{center}
\leavevmode
\hbox{%
\epsfxsize=7.5cm
\epsffile{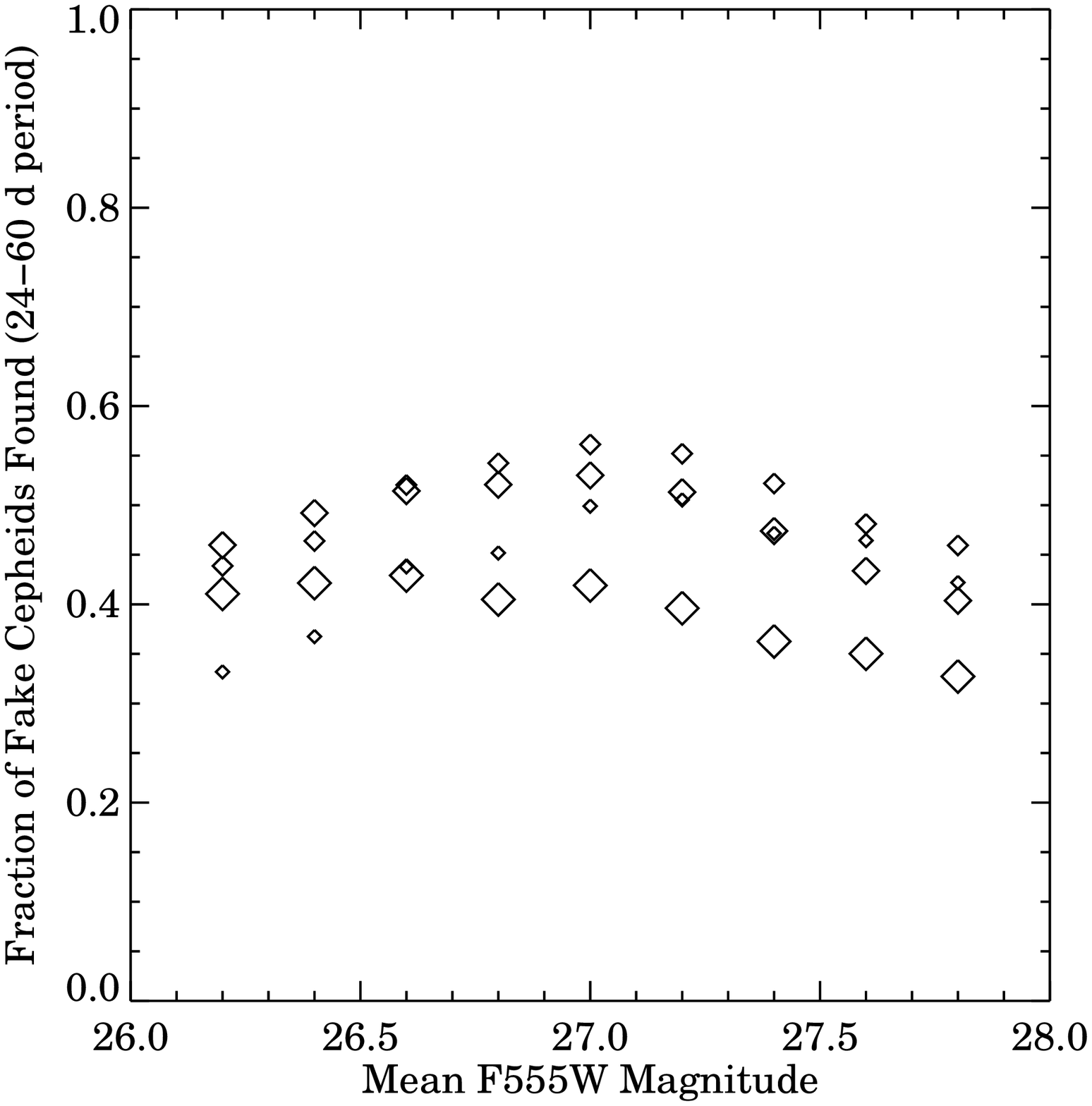}}
\begin{small}
\figcaption{\small Results of simulations for the rate at which our algorithms detect Cepheid variables of a given $F555W$ magnitude.  The data are divided into subgroups according to the periods of the simulated Cepheids
(with the largest symbols used for the average detection rate for
those stars with the longest periods of variation, and the smallest
the shortest).  The dependences of the variable detection rate upon
period and amplitude were complex and nonseparable, requiring us to
interpolate upon a grid of simulation results in our maximum
likelihood analysis.}
\end{small}
\end{center}}
found through a power-law fit to the long-period
tail of the distribution of Large Magellanic Cloud (LMC) Cepheids in
Alcock et al. 1999 to be roughly proportional to the -2.0 power of
period (defining the parameter $\alpha$ used below; i.e., we have
adopted a differential distribution of Cepheids in period of the form
$N(P_R) dP_R \propto P_R^{\alpha} dP_R$).  Even violently changing
this assumption (changing $\alpha$ by $\pm$ 1) led to changes in the
derived distance modulus of less than 0.10 mag.  For the purpose of
this analysis, we adopt the LMC Cepheid Period-Luminosity relations of
Madore \& Freedman (1991) (and, for the likelihood analysis, the
dispersions of LMC Cepheids about that relation) which have been used
by the Key Project on the Extragalactic Distance Scale.

There are two distribution functions required for this
analysis, labelled hereafter as $f_{real}$ and $f_{false}$.  These
represent the probability that a particular star is a real Cepheid
and detected with given properties, or a nonvariable star and
identified as a Cepheid with those properties, respectively.
Based upon the results of our simulations, the former is defined
as a function of observed period $P$, magnitude $m$, and amplitude
$A$, and of the given distance modulus $m-M$ as
\begin{eqnarray}
f_{real}(m,P,A | m-M) \sim p_{detect}(m,P,A) {1 \over 2 \pi \sigma_{m}
\sigma_{P}} \\
\nonumber \times \int_{P_{min}}^{P_{max}}{ e^{-{(P-P_r)^2 \over 2
\sigma_P^2}} e^{-{(m-m_r)^2 \over 2 \sigma_m^2}} P_r^{\alpha}
dP_r},
\end{eqnarray}
where $p_{detect}(m,P,A)$ is the probability of our detecting a
Cepheid that has a given observed magnitude, period, and amplitude,
$P_r$ is the actual, as opposed to
\vbox{%
\begin{center}
\leavevmode
\hbox{%
\epsfxsize=7.5cm
\epsffile{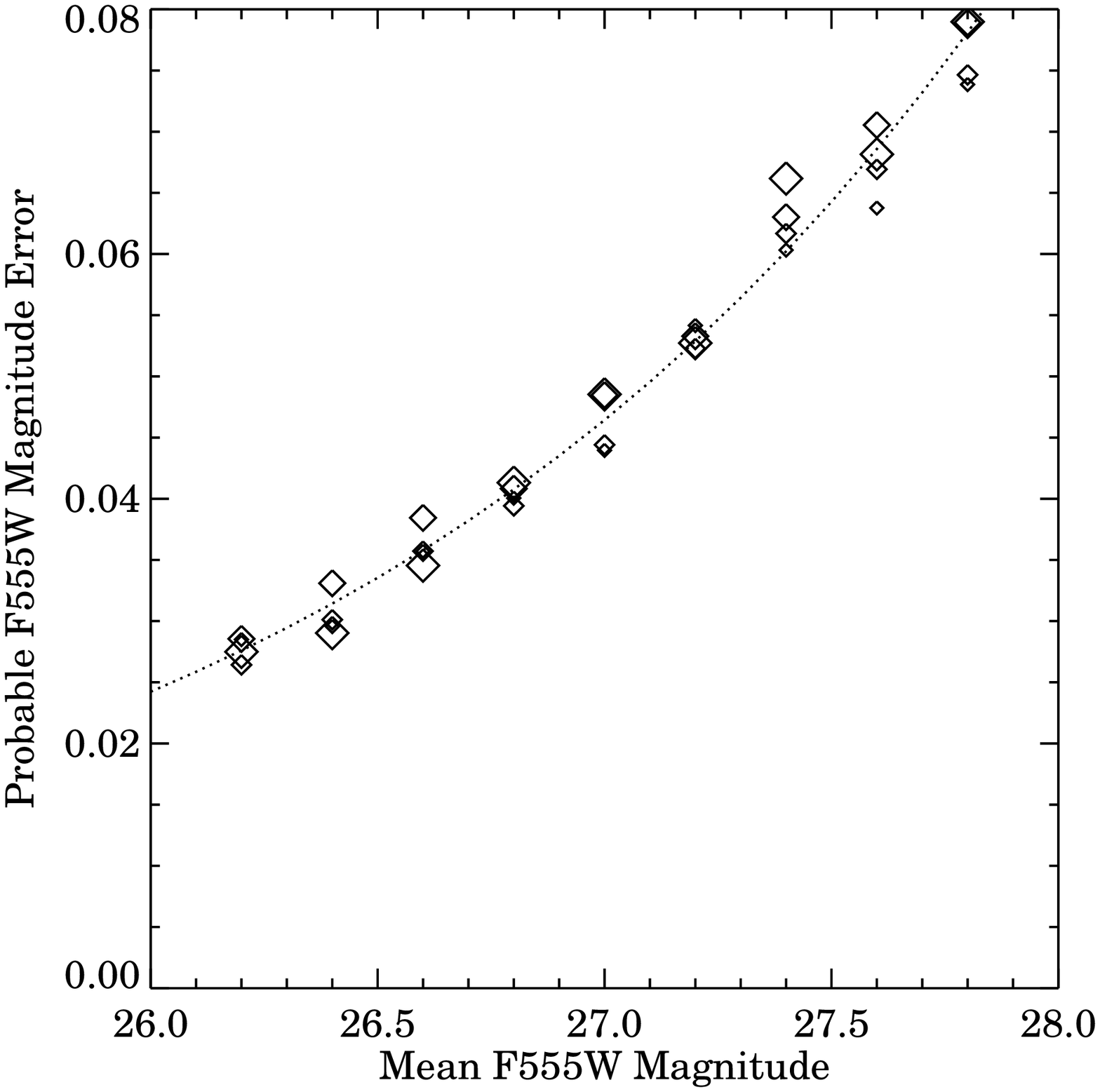}}
\begin{small}
\figcaption{\small Results of simulations for the probable
error in measuring the mean $F555W$ magnitude of a Cepheid as a
function of $F555W$ magnitude.  No trend in this quantity is seen
either with the period or amplitude of the Cepheid's variation; in
this plot, as in Figure 6, symbols of larger size represent errors for
longer-period variables.  In our maximum likelihood analyses, probable
errors are multiplied by an appropriate correction factor, 1.48260, to
yield the corresponding Gaussian $\sigma$.}
\end{small}
\end{center}}
observed, period of a Cepheid,
$m_r$ is the ideal magnitude of a Cepheid for a given distance modulus
and $P_r$ (from the Madore \& Freedman P-L relation), and $\alpha$ is
treated as a constant parameter for the maximum likelihood analysis
describing the distribution in period of actual Cepheids.  In our
Monte Carlo analysis, $\sigma_{P}$ proved to be a complex function of
period, amplitude, and magnitude, while $\sigma_m$ was significantly
dependent only on magnitude (as applied in the maximum likelihood
analysis, $\sigma_m$ has added to it in quadrature contributions from
the dispersion of the P-L relation and estimates of magnitude
measurement errors due to background subtraction and bias correction
uncertainties).  Because of their complicated dependence on all
possible variables, values of $v$ and $\sigma_P$ were obtained by
interpolating within a $9 \times 9 \times 9$ grid in period,
amplitude, and magnitude containing the results of simulations for
these quantities.  The values of $\sigma_m$ were taken from
least-squares fits of empirically chosen functions to the Monte Carlo
results.  Once a distance modulus is chosen, $f_{real}$ is normalized
to make the expectation value of the number of Cepheids existing in our dataset unity:
\begin{equation}
\sum_i{n_i \int{f_{real}(m_i,P | m-M) dP=1}},
\end{equation}
where $m_i$ is the mean magnitude in a bin (0.04 mag wide in our analysis) and $n_i$ is the
number of observed stars with good photometry in that bin. The actual
number
\vbox{%
\begin{center}
\leavevmode
\hbox{%
\epsfxsize=7.5cm
\epsffile{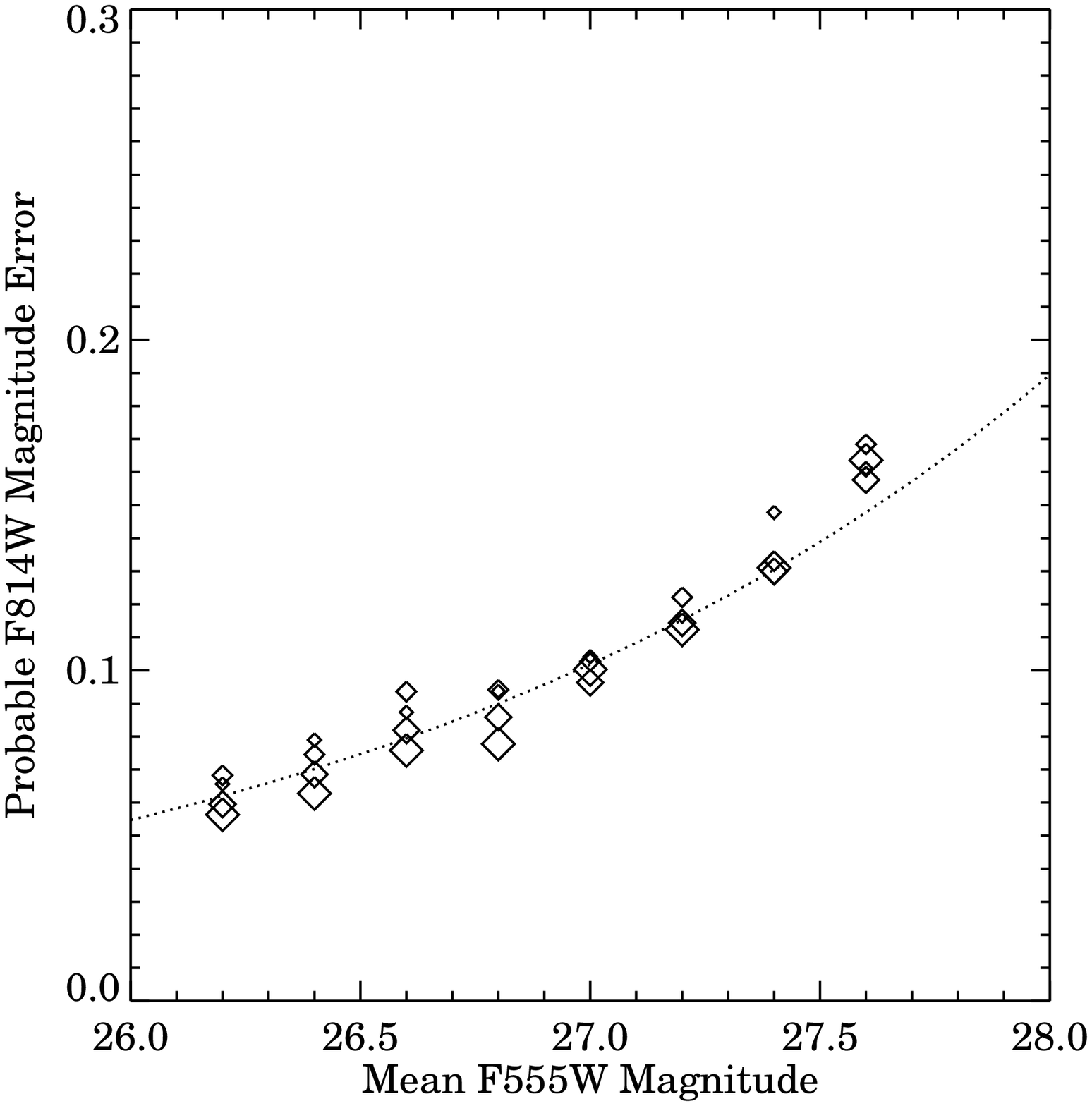}}
\begin{small}
\figcaption{\small Results of simulations for the probable
error in measuring the mean $F814W$ magnitude of a Cepheid as a function of
$F555W$ magnitude.  No trend in this quantity is seen with the
period or amplitude of the Cepheid's variation.}
\end{small}
\end{center}}
of Cepheids in the data will then be a parameter whose value is
determined by the likelihood analysis.  This integral is performed
numerically in 1 day increments over the range of periods accepted for
candidates, 24-60 days.  This normalization may not be perfect,
particularly for the $I$ analysis, so much more significance should be
ascribed to, e.g., results for the difference of the number of
Cepheids from zero than to the exact number of Cepheids found.

The chance that a given unvarying star is selected as a candidate
variable of given properties, $f_{false}$, was found via our
simulations to be proportional to a power law in period and a Gaussian
(of zero mean) in amplitude:
\begin{equation}
f_{false}(m,P,A) \sim {{1} \over{\sqrt{2 \pi \sigma_A}}} e^{-{{A^2}
\over{2\sigma_A^2}}} P^{\beta};
\end{equation}
in our Monte Carlo simulations, $\sigma_A$ proved to be a function of
the magnitude alone and $\beta$ a constant.  Because these parameters
are independent of period and amplitude, $f_{false}$ may be
integrated over these variables analytically.  This distribution was
then normalized such that its integral over the possible periods and
amplitudes for candidates was 1; for an individual candidate, it must
be multiplied by the overall rate of false positives at a given
magnitude, $r(m)$, to yield the probability that that star is a false
positive.

The functional fits to the parameters required by the maximum
likelihood analysis used were:
\begin{equation}
\sigma_{m_V}(m_V)=0.1 \times 10^{-0.2824(28.18-m_V)} $$
$$ \sigma_{m_I}(m_V)= 0.2 \times 10^{-0.2696(28.09-m_V)} $$
$$ \sigma_A(m_V) = \max (0.08,-2.013+0.0791 m_V) $$
$$ \beta = -1.82 $$
$$ r(m_V) = \max (0.05350 \times (m_V-26.6)^{2.714},0),
\end{equation}
where $m_V$ is the $F555W$ magnitude of a given star before
aperture corrections and $m_I$ its corresponding $F814W$
magnitude.

The logarithm of the likelihood is then defined as
\begin{equation}
\ln{\cal{L}}=\sum_{i=1}^{n_{cand}} \ln[ N_{Ceph} f_{real}(m,P,A |
m-M) + r(m) f_{false}(m,P,A)] $$
$$ + \sum_{j=1}^{n_{bin}} n_j \ln\left[ \left(1-N_{Ceph}
f_{real}(m_j | m-M)\right)
\left(1-r(m_j) \right) \right],
\end{equation}
where $n_{cand}$ is the total number of Cepheid candidates, $n_{bin}$
is the number of magnitude bins used, and $N_{Ceph}$ is roughly
equivalent to (and directly proportional to) the number of observed
Cepheids in the dataset, an unknown in the analysis.  The first
summation corresponds to the product (before the logarithm) of the
probabilities that our candidate Cepheids will be detected as such
stars with their given properties; the second, the product of the
probabilities that each of our non-candidate stars are not either
detected Cepheids or false positives.  The logarithm of the
likelihood, and thus the likelihood itself, is maximized over a grid
in the distance modulus $m-M$ and the number of Cepheids in the
dataset $N_{Ceph}$.  Note that for the non-candidates, the
distribution functions have been integrated over period and amplitude.

We have tested our maximum likelihood techniques by applying them to
datasets containing both nonvariable stars (potentially false
positives) and a set of simulated Cepheids with a fixed distance
modulus and a realistic distribution of properties. If the number of
real Cepheids was large enough and NGC~4603 placed near enough that
\vspace{.33 in}
\vbox{%
\begin{center}
\leavevmode
\hbox{%
\epsfxsize=7.5cm
\epsffile{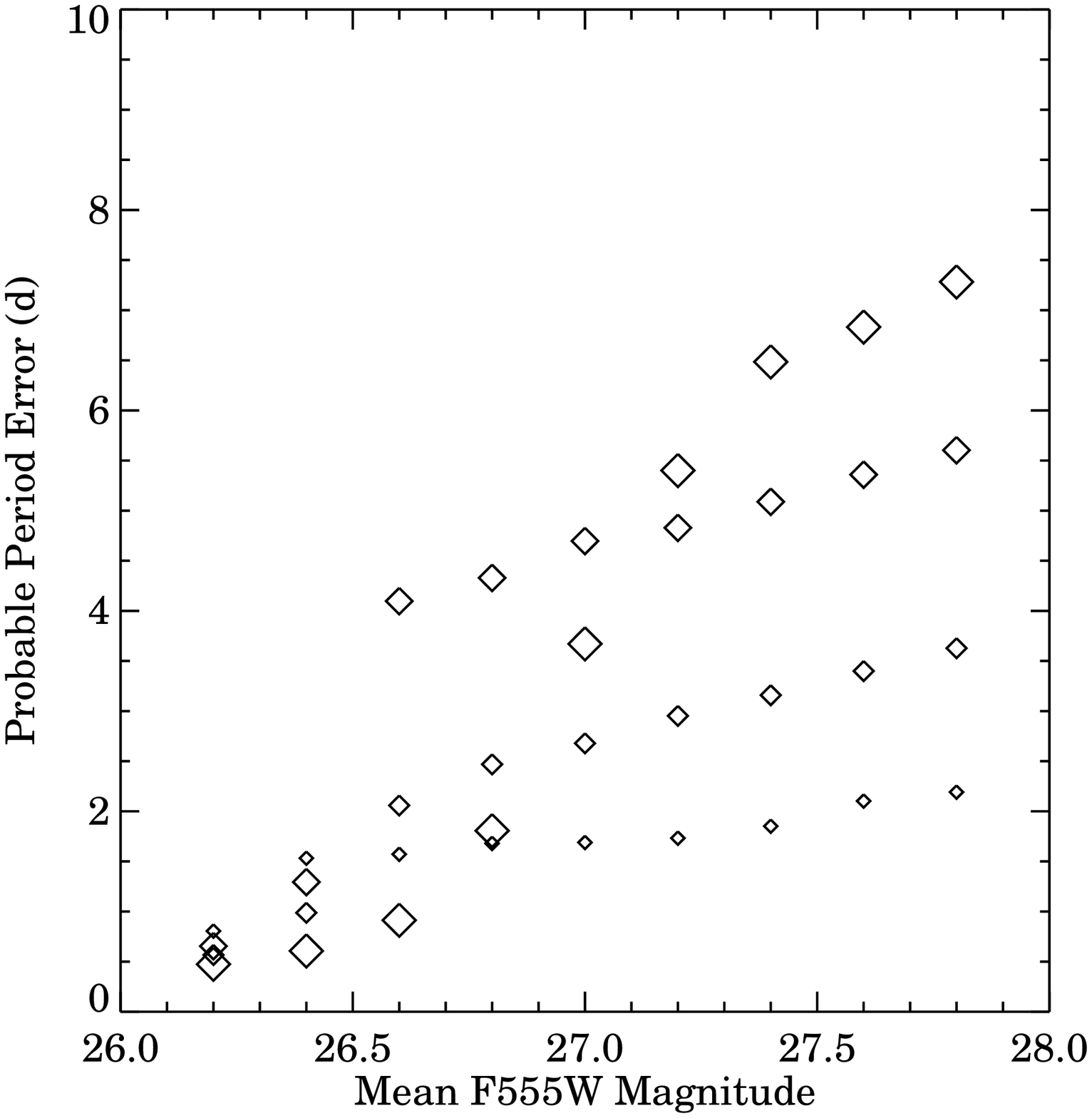}}
\begin{small}
\figcaption{\small Results of simulations for the probable
error in measuring the period of a Cepheid as a function of magnitude.
Like the Cepheid detection rate, this quantity proved to be dependent
on period and amplitude in a complex, nonseparable fashion; therefore,
interpolations of the results of simulations were used for it in our
maximum likelihood analysis.  Due to aliasing, the longest-period
Cepheids have the greatest errors in period determination (as can be
seen in Figure 1).}
\end{small}
\end{center}}
a significant number of the input Cepheids are found by the
variable search procedure, then our techniques effectively recovered
the input distance.  If those conditions are not met, the
highest-likelihood solutions generally prove to be those in which the
number of Cepheids in the dataset is minimized or the distance modulus
is maximized, i.e., cases in which the chance of observing a Cepheid
would be as small as possible.  Such solutions are easy to recognize
and did not occur in our analysis of NGC~4603.
 
This analysis was performed independently using the $V$ and $I$ mean
magnitudes of our candidates to determine the distance modulus.  From
the $V$ analysis, we determine that the hypothesis of no Cepheids
present is excluded at $> 9\sigma$, that $43 \pm~7$ actual Cepheids
are present in our dataset, and that NGC~4603 has a distance modulus
of $33.15_{-0.10}^{+0.11}$ (1$\sigma$ random errors) before correction
for metallicity and dust extinction.  The $I$ analysis yields a poorer
constraint, with a Cepheid signal present at $> 7 \sigma$ and an
uncorrected distance modulus measurement of $32.97_{-0.09}^{+0.15}$.
See Figures 11 and 12 for plots of the resulting likelihood contours.
The location of our candidates in the NGC~4603 color-magnitude diagram
is shown in Figure 13.  Figures 14 and 15 illustrate the differences
between the distributions of candidate Cepheids in magnitude and color
and those expected for false positives.  The excess candidates beyond
the false positives do seem limited in their brightness and colors in
the fashion expected for Cepheids of a variety of periods and
reddenings.

Since we have obtained substantial knowledge about the distribution in
properties of real Cepheids and false positives through our maximum
likelihood analysis, we can estimate the probability that a given
candidate is in fact a Cepheid; the resulting probabilities are listed
in Table 2.  $V$ and $I$ period-magnitude plots for the potential
Cepheids we have found in NGC~4603 (containing essentially the
\vspace{.13in}
\setcounter{figure}{10}
\vbox{%
\begin{center}
\leavevmode
\hbox{%
\epsfxsize=7.5cm
\epsffile{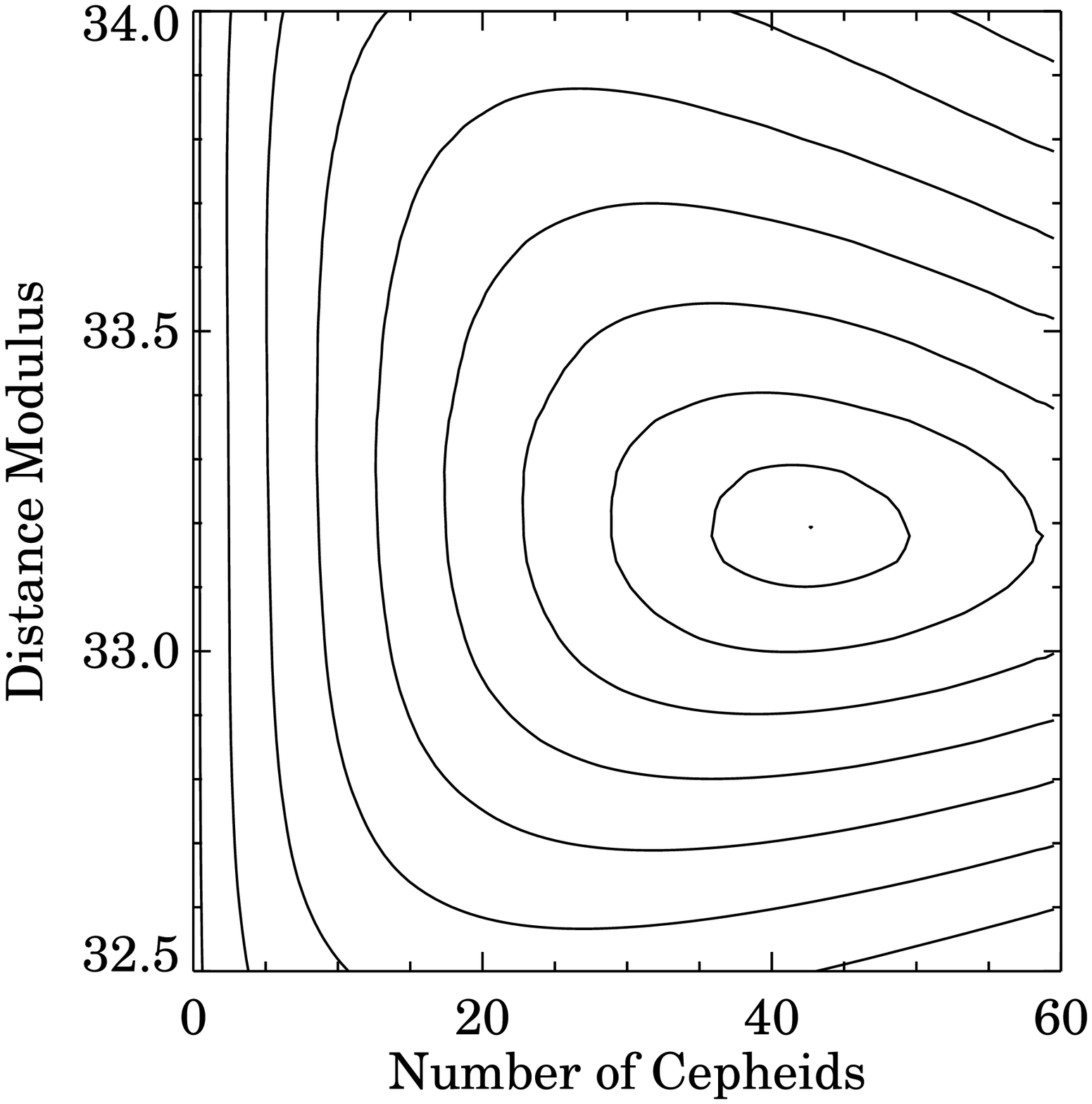}}
\begin{small}
\figcaption{\small Results of our maximum-likelihood analysis
using $V$ mean magnitudes of candidate Cepheids to determine the
distance to NGC~4603.  The contours represent 1,2,3,4, etc. $\sigma$
limits on the measured parameters.  We confirm that Cepheids are
present in our data set at $>9 \sigma$.}
\end{small}
\end{center}}
same
information) may be found in Figures 16 and 17.  Those candidates
found to have greater than 50 \% probability of being Cepheids in both
the $V$ and $I$ maximum likelihood analyses have their
simulation-based error bars (as used in the analyses) depicted on the
plots.  

Such higher-probability candidates may be used to provide an
illustration of the workings of our maximum likelihood procedure.
These stars should have a relatively high value of $f_{real}$, so they
must agree with the expected magnitude of a Cepheid of the same
measured period given our choice of distance modulus within the
estimated errors.  However, they should also have a relatively small
value of $r(m) f_{false}$.  Given the strong magnitude dependence, we
expect such stars to be brighter than the typical candidate.  Thus, if
we were to calculate the mean distance modulus predicted from the
properties of such stars, we would expect it to be fairly consistent
with but biased low compared to that obtained from the full maximum
likelihood analysis.  This is borne out by such a calculation for,
e.g., those stars that have $>80\%$ probability in both the $V$ and
$I$ analyses; they give a value of $32.95 \pm~0.10$ for the $V$
modulus and $32.80 \pm~0.08$ for $I$, 0.20 and 0.17 mag less than
those obtained from the full procedure.  The maximum likelihood
technique does not simply determine a distance modulus weighting stars
according to their probability of being Cepheids, but instead
incorporates as much information as possible about how effectively we
can find such stars, minimizing incompleteness/Malmquist-type biases.

\subsection{Uncertainties and Corrections}

In addition to the statistical uncertainties in our measurements of
the distance modulus of NGC~4603, which
\vbox{%
\begin{center}
\leavevmode
\hbox{%
\epsfxsize=7.5cm
\epsffile{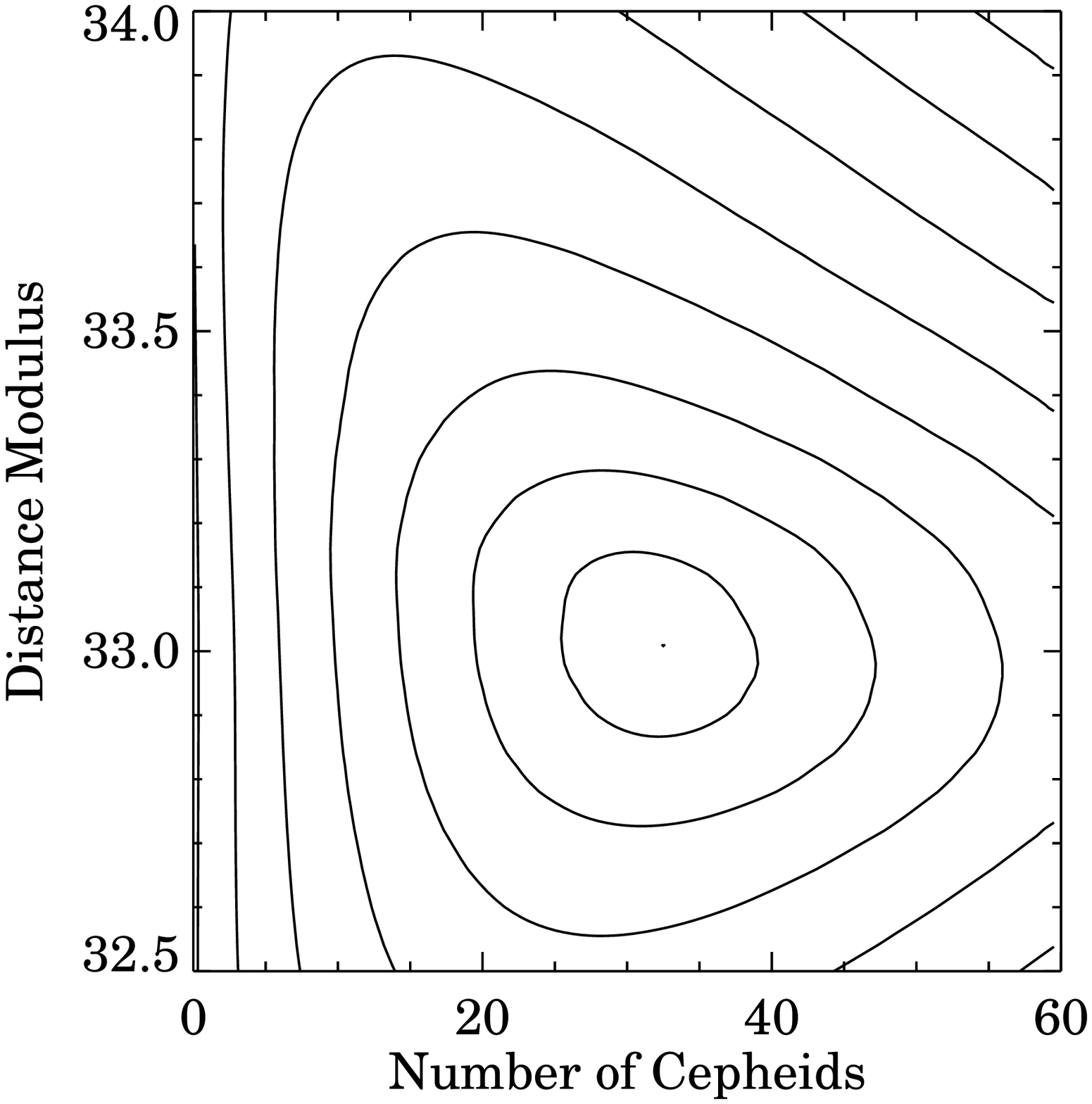}}
\begin{small}
\figcaption{\small Results of our maximum-likelihood analysis
using $I$ mean magnitudes of candidate Cepheids to determine the
distance to NGC~4603.  The contours represent 1,2,3,4, etc. $\sigma$
limits on the measured parameters.  We confirm the detection of
Cepheids in the $I$ band at $>7 \sigma$.}
\end{small}
\end{center}}
\vbox{%
\begin{center}
\leavevmode
\hbox{%
\epsfxsize=7.5cm
\epsffile{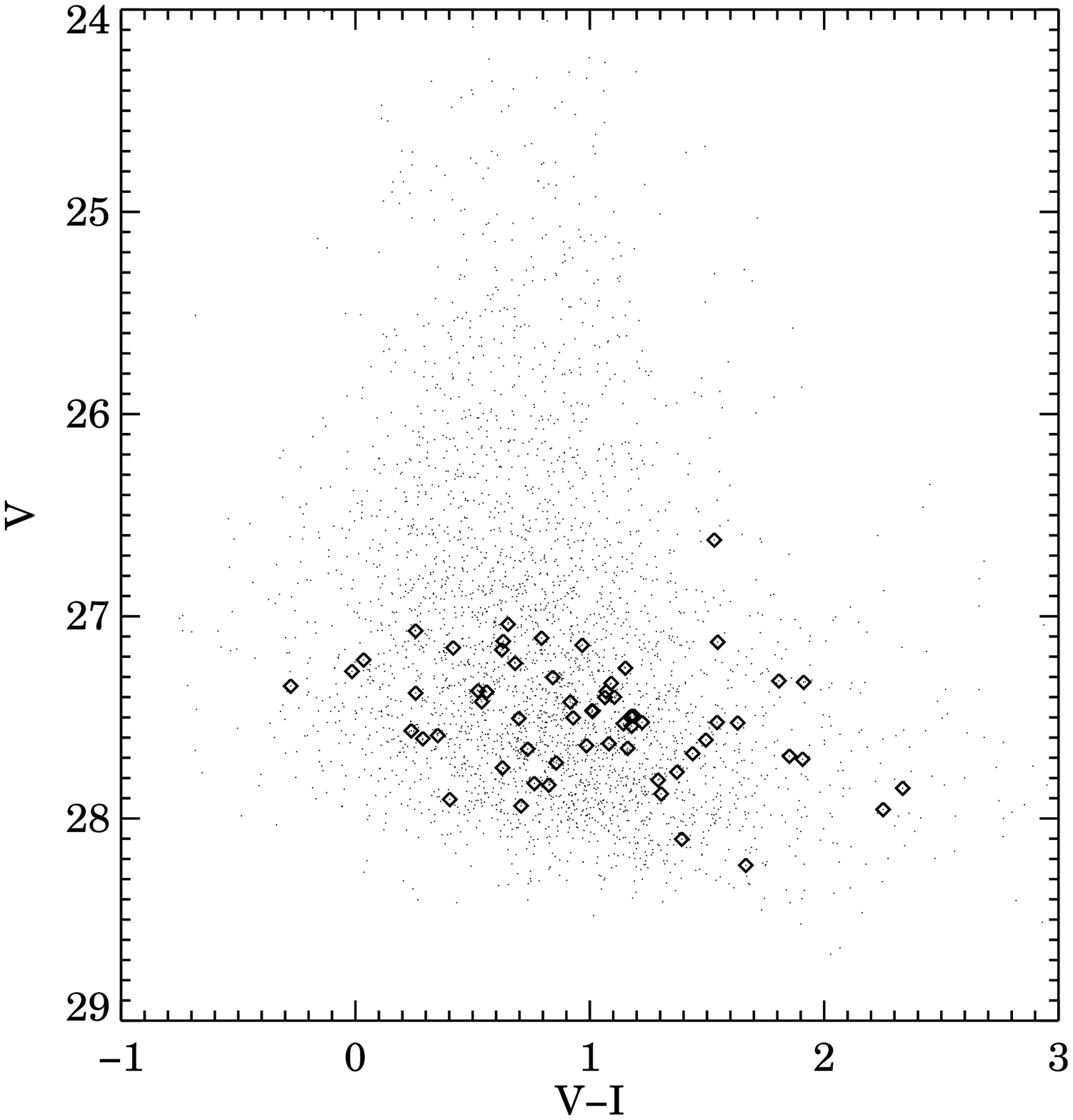}}
\begin{small}
\figcaption{\small A color-magnitude diagram for stars on Chip 1.  Candidate
Cepheids are indicated by open symbols.}
\end{small}
\end{center}}
were determined by our maximum
likelihood analysis, our results are also subject to a number of
potential sources of systematic error.  In this subsection, we will
attempt to estimate the amounts of possible error due to the
calibration of photometry and our analysis techniques, to
uncertainties in the Cepheid P-L relation calibration, and to our
limited knowledge of physical conditions in and towards NGC~4603, and
make whatever well-established corrections possible to our distance
moduli.

Uncertainties in the HST zero point of $\pm 0.05$ magnitudes in $V$ or
$I$ affect our measurements of the NGC 4603 distance in much the same
fashion as Key Project distances (Hill et al. 1998), with one
important difference: because the mean magnitudes of a given star are
not as well determined, we are unable to use a measurement of $E(V-I)$
to measure reddening, so relative zero point errors do not propagate
into our results as they do in the Key Project methodology.  A mean
difference of 0.08 mag in $V$ between DoPHOT and non-bias-corrected
ALLFRAME photometry for our candidates was found.  This may be due to
differences in the characteristics of any biases that occur when
averaging ALLFRAME and DoPHOT results at these faint magnitudes.  We
adopt the bias-corrected ALLFRAME results here and include half this
difference as a potential systematic error in $V$ magnitudes.  In the
absence of sufficient DoPHOT comparison photometry in $I$, we consider
a 0.10 mag systematic error to be possible, though substantially
larger than any found in prior studies.

 We also must consider
uncertainties in the distance modulus resulting from the maximum
likelihood methodology and the Monte Carlo fits that were used to
define $f_{real}$ and $f_{false}$.  Changing the assumed false
positive rate radically (e.g., by 50\%) altered the resulting distance
modulus constraints by at most 0.09 mag in both $V$ and $I$.
Considering also the differences in measured distance modulus
\vbox{%
\begin{center}
\leavevmode
\hbox{%
\epsfxsize=8.9cm
\epsffile{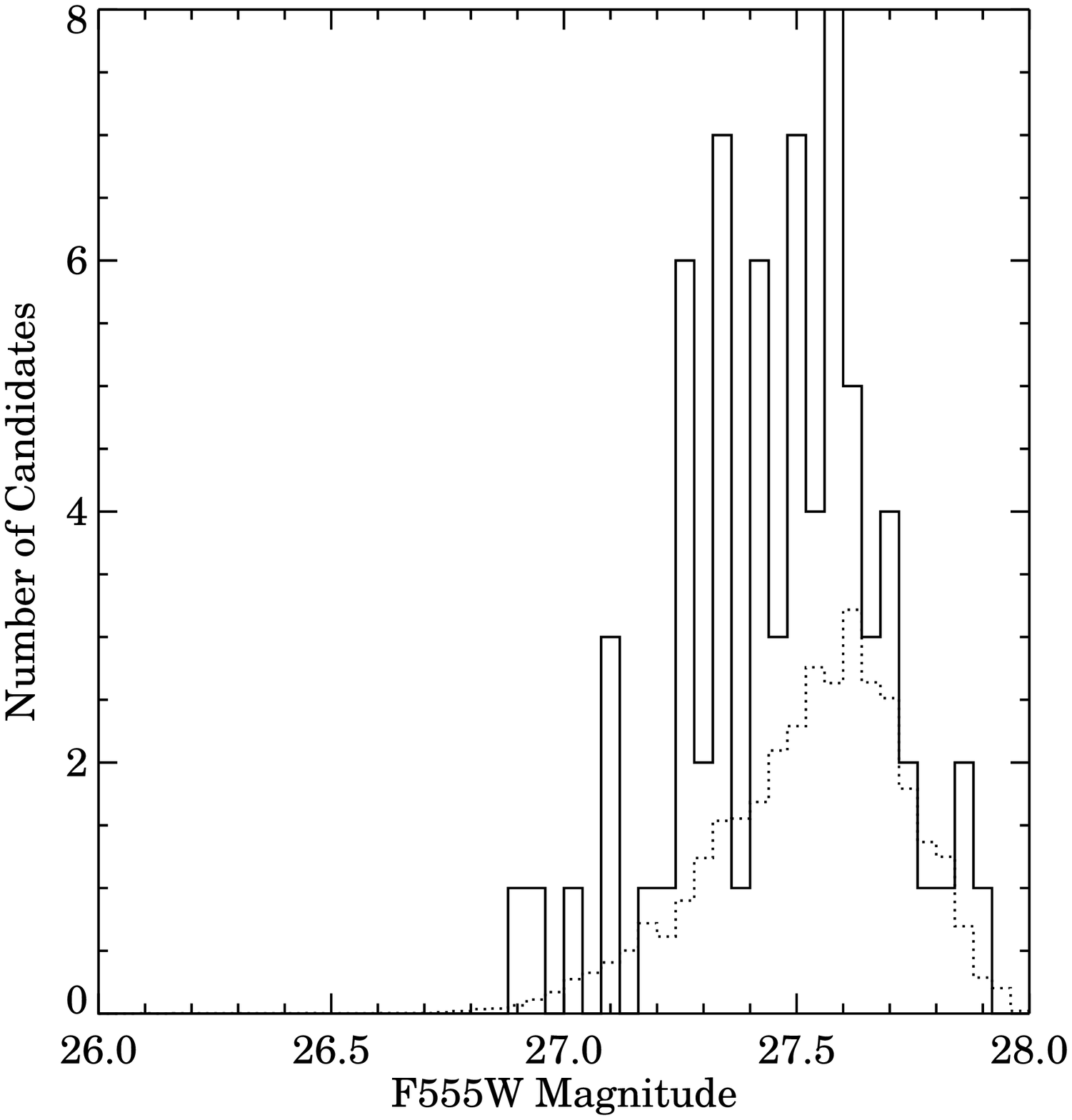}}
\begin{small}
\figcaption{\small Histogram of the $F555W$ magnitude
distribution of candidate Cepheids (solid line) and that expected for
false positives given the distribution in magnitude of observed stars
(dashed line).}
\end{small}
\end{center}}
exhibited when the power-law parameter for the distribution in period
of real Cepheids, $\alpha$, is changed by $\pm 1$, we find potential
systematic errors in the maximum likelihood procedure of 0.14 mag for
$V$ and 0.12 mag for $I$. If we add the corresponding errors in
quadrature, we find that systematic errors in photometry and in our
analysis techniques should be less than 0.15 mag in $V$ and 0.16 mag
in $I$.  Because we account for the lower probability of detecting
faint Cepheids via our maximum likelihood technique and fix the slope
of the P-L relation used, the effects of incompleteness bias should be
minimal.

Our results are also subject to possible systematic errors in the
calibration of the Cepheid P-L relation.  Indeed, one of the largest
remaining systematic uncertainties in the extragalactic distance scale
is our limited knowledge of the distance to the Large Magellanic
Cloud, which currently provides the fiducial standard Cepheid
calibration.  For the Key Project, this uncertainty has been taken to
be $\pm$ 0.13 mag; we adopt this value so that if the distance to the
LMC is better determined in the future our distance determination may
be easily adjusted in concert with theirs.  We also adopt the Key
Project's estimate of potential errors within the LMC $V$ and $I$ P-L
calibrations of $\pm 0.05$ magnitudes (see, e.g. Rawson et al. 1997).

A number of potential systematic errors in our distance modulus could
be the result of physical effects.  First, as an Sc galaxy or,
alternatively, one with maximum circular velocity of 220 \kms\
(Giovanelli et al. 1997), we may expect Cepheids in NGC~4603 to
possess substantially higher 
\vbox{%
\begin{center}
\leavevmode
\hbox{%
\epsfxsize=8.9cm
\epsffile{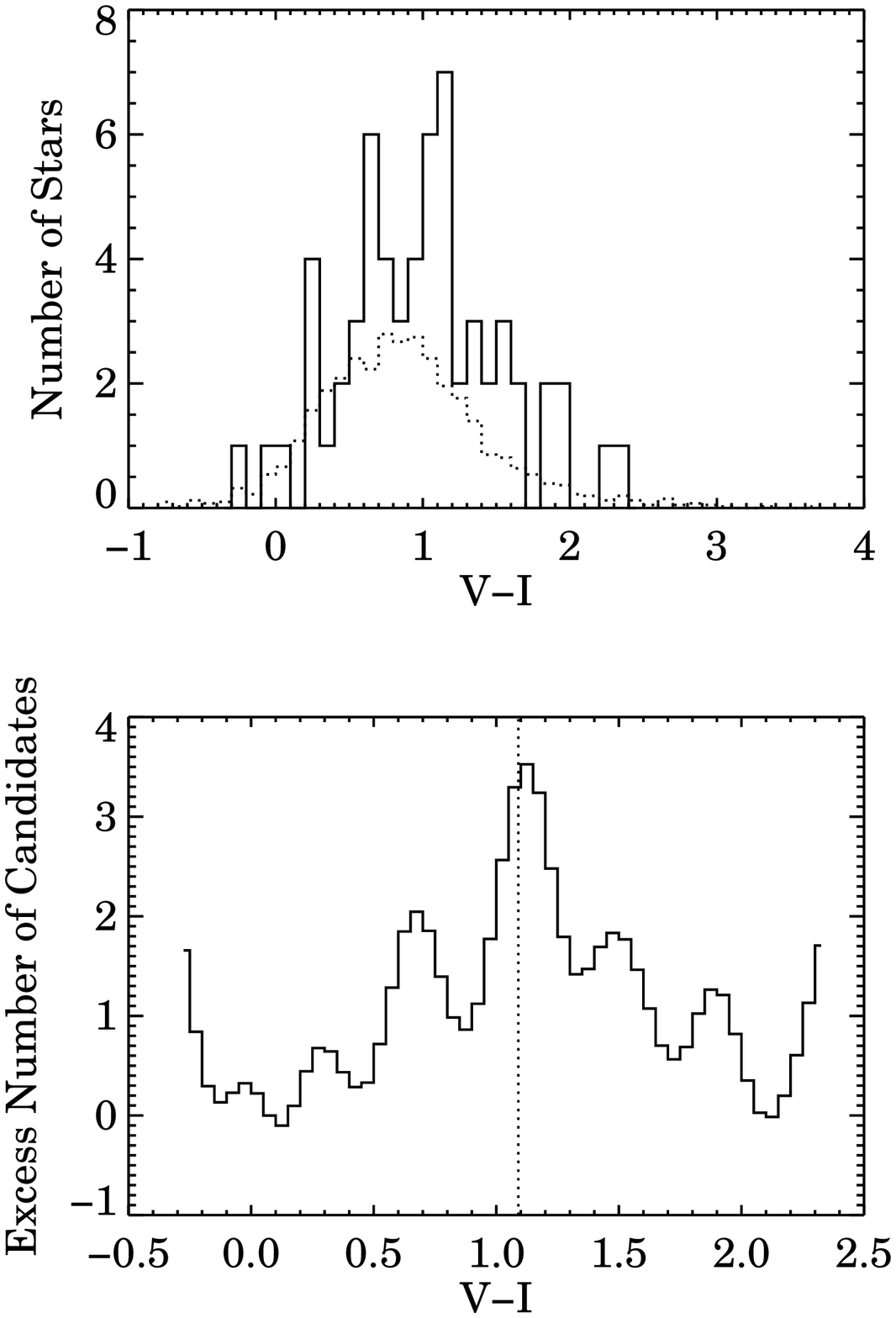}}
\begin{small}
\figcaption{\small (upper panel) Histogram of the $V-I$ colors of
candidate Cepheids (solid line) and that expected for false positives
given the color distribution of observed stars (dashed line).  (lower
panel) The difference between the two histograms, smoothed with a
Gaussian kernel.  The dotted line indicates the typical color expected
for a Cepheid of period 35 d reddened by foreground Galactic
dust as measured by Schlegel, Finkbeiner, \& Davis (1998).  The
distribution appears very consistent with Cepheids of a range of
reddenings, with no similar excess of blue stars.}
\end{small}
\end{center}}
metallicity than those in the LMC, by
roughly 0.40 $\pm 0.20$ dex at the radius of the PC field (applying
the results of Zaritsky, Kennicutt, and Huchra 1994 to obtain values
for the typical metallicity and metallicity gradient in NGC~4603).
Using the relation of Kennicutt et al. (1998), we should therefore
expect that our distance modulus is an underestimate by 0.096 $\pm
0.081$ mag.  It should be noted that other studies have found larger,
but still statistically equivalent given the error bars, metallicity
effects (Sasselov et al. 1997, Kochanek 1997, Nevalainen \& Roos
1998), while theoretical calculations predict effects that are minimal
or opposite in sign (Alibert et al. 1999, Musella 1999).  We therefore
make no correction, and consider the entire 0.096 mag to be a
potential systematic error.

Another potential physical effect is extinction by dust along the
line-of-sight to the Cepheids, either within NGC~4603 or our own
Galaxy. Unfortunately, the Centaurus cluster lies behind a region
where substantial emission from Galactic dust has been observed; the
effect of this dust should therefore be quite appreciable.  We thus
must correct the distance moduli we have found for Galactic foreground
dust absorption of $A_V$=0.54 $\pm 0.08$ magnitudes ($A_I$=0.33
magnitudes, using a typical Galactic extinction law), taken from the
extinction map of Schlegel, Finkbeiner \& Davis (1998).  After
correction for foreground extinction, our data yield $E(V-I)_{internal}=-0.04^{+0.14}_{-0.18}$ (random); we constrain the reddening
due to dust within NGC~4603 only poorly.  To place limits on its
effects, we may safely assume that internal dust will yield
E($V-I$)$\geq$ 0, of course; an examination of Key Project papers
studying galaxies of similar inclinations indicates that E(V-I) $<
0.07$ due to internal reddening is also a reasonable
assumption. Conversion to $A_V$ with a typical Galactic extinction law
indicates that we might therefore expect that our $V$ distance modulus
should be reduced by as much as 0.17 mag in correcting for extinction
by dust within NGC~4603, and our $I$ modulus by as much as 0.10 mag.
We thus adopt -0.09 magnitude as an estimate of the possible 1$\sigma$
systematic error from internal extinction in $V$, and -0.05 mag in
$I$.

To estimate the total potential systematic errors, we add possible
errors from physical effects in quadrature to the systematic
uncertainties of the modulus from our photometric and maximum
likelihood techniques and that from the P-L relation calibration,
yielding total systematic 
\vbox{%
\begin{center}
\leavevmode
\hbox{%
\epsfxsize=8.9cm
\epsffile{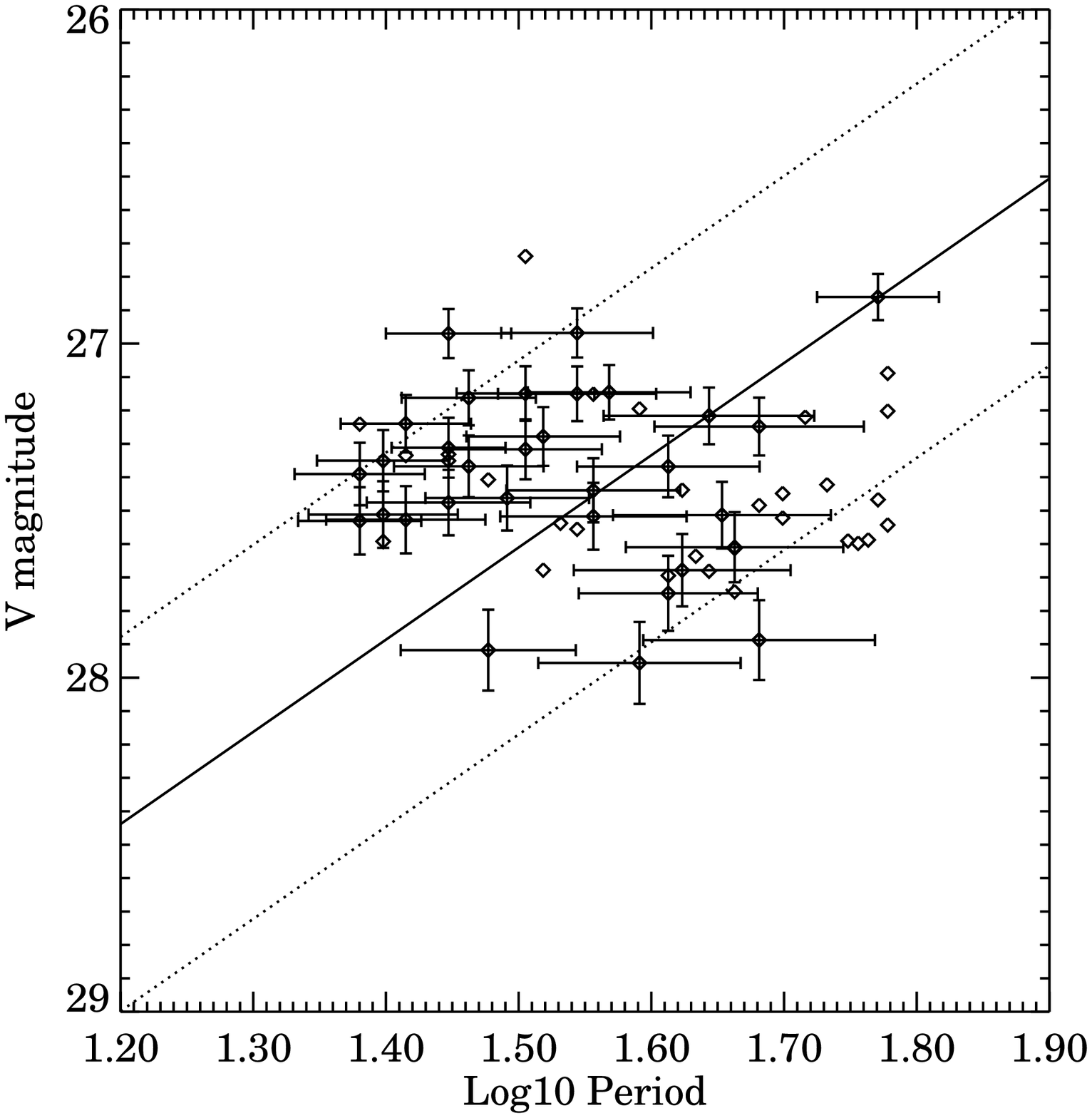}}
\begin{small}
\figcaption{\small The $V$ P-L relation for our candidate
Cepheids.  The solid line depicts the LMC P-L relation shifted to
the distance modulus we have obtained; dotted lines indicate the
$2-\sigma$ scatter of LMC Cepheids about that relation.  Those candidates
with more than 50\% probability of being Cepheids in both the $V$ and $I$
analyses are plotted with the error bars used for them in
the maximum likelihood analysis (drawn from our simulations).}
\end{small}
\end{center}}
\vbox{%
\begin{center}
\leavevmode
\hbox{%
\epsfxsize=8.9cm
\epsffile{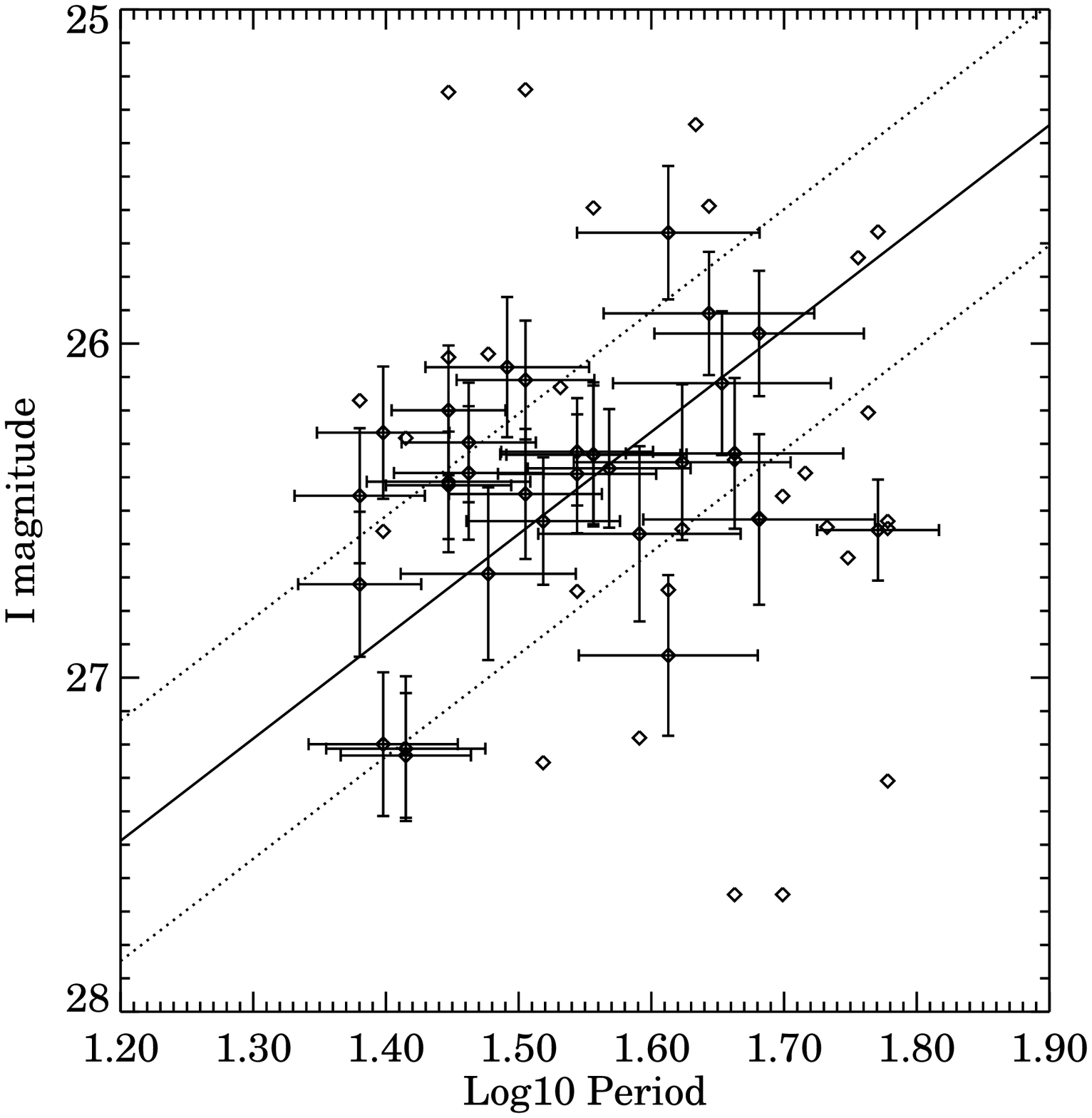}}
\begin{small}
\figcaption{\small The $I$ P-L relation for our candidate
Cepheids.  The solid line depicts the LMC P-L relation shifted to the
distance modulus we have obtained; dotted lines indicate the
$2-\sigma$ scatter of LMC Cepheids about that relation.  Those
candidates with more than 50\% probability of being Cepheids in both
the $V$ and $I$ analyses are plotted with the error bars used for them
in the maximum likelihood analysis (drawn from our simulations).}
\end{small}
\end{center}}
uncertainties of +0.23/-0.24 magnitudes in
$V$ or $\pm 0.23$ magnitudes in $I$. Combining all effects, we thus
find from the $V$ analysis that NGC~4603 has a distance modulus of
$32.61_{- 0.10}^{+0.11}$ (random, 1 $\sigma$) $^{+0.24}_{-0.25}$
(systematic), corresponding to a distance of $33.3^{+1.7}_{-1.5}$
(random, 1 $\sigma$) $^{+3.8}_{-3.7}$ (systematic) Mpc.  The $I$
analysis provides a quite consistent result, yielding a distance
modulus of $32.65_{-0.09}^{+0.15}$ (random, 1 $\sigma$) $\pm 0.24$
(systematic).

\subsection{Implications}

Previous studies have obtained widely differing measurements of the
distances to, and hence peculiar velocities of, Cen30 and Cen45, even
when using the same techniques.  Larger $D_n - \sigma$ samples
(e.g. that used in the Mark III catalog [Willick et al. 1997], which
includes 22 galaxies in Cen30 and 9 in Cen45, as opposed to 9 and 4,
respectively, in Faber et al.), for instance, have placed Cen30 as far
as or even $behind$ Cen45, with peculiar velocities of -110 \kms\ and
+1515 \kms\, respectively, in great contrast to the earlier results.
The hypothesis that Cen30 and Cen45 may lie at the same distance was
first advanced by Lucey, Currie and Dickens (1986a) on the basis of a
number of relative distance measures.  It appears from some studies as
though the Centaurus cluster may be in the midst of a substantial
merger, with Cen45 falling into Cen30 and acquiring a rapid velocity
thereby.  On the other hand, some distance measurements, particularly
those considered by Lynden-Bell et al. (1988), imply that Cen30 and Cen45
moving rapidly relative to the Local Group.

A flow of the center of mass of the Cen30/Cen45 system with 
such high speed -- faster than the motion of the Local Group itself --
would suggest the existence of a substantial attracting mass.  From the
results of the first $D_n-\sigma$ studies, Lynden-Bell et al. (1988)
hypothesized the existence of a ``Great Attractor,'' a large
concentration of matter lying beyond the Centaurus cluster.  However,
neither optically nor $IRAS$-selected samples of galaxies have revealed
regions of overdensity sufficient to explain these motions.  In fact,
according to redshift surveys, the Centaurus cluster itself, when
combined with Hydra and Pavo-Indus-Telescopium on the other side of
the galactic plane, should constitute the major local attractive
point.  Centaurus should therefore be approximately at rest in the
Cosmic Microwave Background reference frame, and the bulk of the
motion of the Local Group driven by its mass overdensity.  In the
Local Group frame one then expects to observe negative peculiar
velocities in the direction of Centaurus, as there is expected to be a
strong reflex dipole pattern from the motion of the Local Group
itself.  One possible explanation for the discrepancy between
predicted and observed flows in this region has been provided by
Guzman and Lucey (1993), who have suggested that $D_n - \sigma$
distances can be compromised by age effects and that the large outflow
of Centaurus is potentially suspect as a result; however, there is no
particular reason to expect that galaxies in the Centaurus region
should be younger than others in our neighborhood.  If there is in fact
only a very weak reflex signature in the velocity of the Centaurus
cluster, the density parameter of the Universe must be very low, too
low to explain the infall pattern around the Virgo supercluster.  

The Cepheid distance measurement we have obtained may be used to set
limits on such a flow.  Our result is most easily compared to studies
using other distance indicators and a peculiar velocity determination is
most straightforwardly made by converting to velocity distance.  This
may be accomplished by multiplying the distance obtained by an
appropriate value for Hubble's Constant based upon the same
calibration; we use the Key Project's most recent estimate for
Hubble's Constant based upon Cepheid data analyzed similarly to that
presented here, 72 $\pm 5$ (random) $\pm 12$ (systematic) \kms\
$\mathrm{Mpc}^{-1}$ (Madore et al. 1999).  We then determine a velocity
distance for NGC~4603 of 2395 $\pm 306$ (random) $\pm 281$
(systematic) \kms.  Note that this value should not be altered by any
recalibrations of the zero point of the Cepheid distance scale because
our distance measurements and those of the Key Project would be
affected in the same way.

A variety of other estimates of the velocity distance of of NGC~4603
and of Cen30 as a whole are presented in Table 3.  To allow more
effective comparison, the presumed velocity of Cen30 in the Local
Group frame(``$cz_{Cen30,LG}$'') and number of galaxies included in
each study are also listed; each of the $D_n-\sigma$ samples in the table
includes the preceding work as a subset.

Our result agrees well with estimates of the distance to Cen30 based
on global analyses of the properties of cluster galaxies.  Jerjen \&
Tammann (1997), for instance, find from an analysis of galaxy
luminosity functions that Cen30 is 1.63 $\pm$ 0.15 mag beyond Virgo.
Taking the Virgo distance modulus to be 31.07 $\pm$ 0.07 (random; from
Freedman et al. 1998, excluding the 3.9$\sigma$ outlier NGC 4639 from
the average), this yields distance modulus 32.70 $\pm$ 0.16 (random).
Studies of surface brightness fluctuations, too, find distances
consistent with that we have obtained for NGC~4603; recent results of
Tonry et al. (1999) yield a mean distance of $2524 \pm 435$ (random)
for 8 galaxies in Cen30.

Our result may also be compared to peculiar velocity predictions based
on the gravity field inferred from the full sky $IRAS$ survey or from
surveys of optically selected galaxies (Nusser and Davis 1995).  For
example, using the gravity field derived for the $IRAS$ 1.2 Jy survey
and assuming $\beta = 0.5$ leads to a predicted peculiar velocity
$v_p$ (in the LG frame) of 14 \kms\ versus a Cepheid inferred $v_p$ of
-74 $\pm 306$ (random) \kms\ if the redshift of NGC~4603 is left at
its observed value, $cz_{lg} = 2321$ \kms, which should be appropriate
if it is in fact a field galaxy.  If we instead compare to the central
redshift of Cen30, $\approx 2807$ \kms\ in the Local Group frame
(Lucey et al. 1986a), then its predicted $v_p$ is -90 \kms, while the
Cepheid distance would imply $v_p = 412$ \kms.  The predicted and
observed peculiar velocities disagree in this case by $\sim 1.2
\sigma$.

Since we only have been able to perform a Cepheid distance analysis
for one galaxy, we cannot claim to have established unambiguously the
distance to the Centaurus cluster; while some studies have included
NGC~4603 in Cen30, for instance, others have not. Indeed, as
illustrated in Table 3, the location of Cen30 itself in redshift
space, not only real space, has varied substantially from analysis to
analysis, reflecting in no small part the large velocity dispersion
and limited numbers of cluster spirals (Stein et al. 1997).  It is
worthy of consideration, though, that those studies that do exclude
this galaxy from Cen30 place it nearer to us than the cluster itself
(as in Willick et al. 1997, though the groupings used for Mark III
spirals tend to place Cen30 at a substantially higher velocity than
that found in other studies), lending support to the higher distance
estimates for the cluster.  At worst, our distance measurement should
provide a lower limit on the distance to Cen30, and hence an upper
limit on its peculiar velocity.  

Our results are most easily reconciled with those of recent
velocity-distance calibrated studies if NGC~4603 is treated as an
object in the foreground of the Centaurus cluster.  That is a rather
reasonable scenario; previous studies (Bernstein et al. 1994, Willick
et al. 1995, Willick 1999) have found that the Tully-Fisher distances
of what are nominally cluster spirals correlate well with their
(rather than their clusters') redshifts.  As a galaxy with a
Tully-Fisher distance, NGC~4603 should be subject to the same
selection effects.  We note that the velocity distances determined
from the two largest samples of Cen30 galaxies listed in Table 3 are
in excellent agreement with each other though those distances were
obtained via different methods and calibrated separately, and those
measurements agree well with the $IRAS$ 1.2 Jy survey-predicted peculiar
velocity for Cen30.  Our distance measurement for NGC~4603 is in good
accord with a variety of Tully-Fisher measurements of the distance for
that galaxy but agrees more poorly with the best measurements of the
distance to Cen30 as a whole.  Under very reasonable assumptions, we
may conclude that Tully-Fisher distances, and therefore (based on
their consistency for Cen30) those obtained via the $D_n-\sigma$
technique as well, agree with the Cepheid distance scale and
$IRAS$-predicted peculiar velocities to at least as far away as the
Centaurus cluster.

The rough agreement of the results of this analysis with other studies
of the Centaurus cluster is encouraging.  For a firmly established
Cepheid distance to Centaurus, a similar study would have to be
performed on more galaxies, preferably including ones that show more
definitive evidence of location in the cluster core (e.g. stripping of
galactic gas) ensuring that members of Cen30 are observed.  However,
the substantial resources in HST time required with current
instrumentation and the extremely extensive data analysis effort
needed to produce a convincing result means that such efforts should
most likely await the installation of the Advanced Camera for
Surveys. Finding Cepheids at the distance of the Centaurus cluster is
currently possible, but difficult indeed.

-----------------

\acknowledgments
\centerline {\bf Acknowledgements}

We would like to thank Jay Anderson for his efforts to provide
secondary photometry for this study and Jeff Willick for his
assistance in interpreting Mark III data.  We also would like to thank
our program coordinators at STsCI, Doug van Orsow and Christian Ready,
for their assistance, and the anonymous referee for helpful comments.
This work was supported by NASA grants GO-06439 and GO-07507 from the
Space Telescope Science Institute (operated by AURA, Inc. under NASA
contract NAS 5-26555).  J. A. N. acknowledges the support of a
National Science Foundation Fellowship and the Berkeley Fellowship.

\begin{deluxetable}{ccccc}
\tablewidth{31pc}
\tablecaption{Journal of Observations } 
\tablehead{
\tablevspace {-0.25em}  \multicolumn{1}{c}{Mean Heliocentric Julian
Date} & \multicolumn{1}{c}{UT Date} & \multicolumn{1}{c}{Filter} & \multicolumn{1}{c}{Exposure
Time (s)} & \nl
\tablevspace {-0.75em}
}
\startdata
2450230.548 & 26 May 1996 & F555W & 7400 \nl
2450235.439 & 31 May 1996 & F555W & 7400 \nl
2450242.340 & 7 June 1996 & F555W & 7400 \nl
2450248.168 & 13 June 1996 & F555W & 7400 \nl
2450255.068 & 20 June 1996 & F555W & 7400 \nl
2450262.909 & 27 June 1996 & F555W & 7400 \nl
2450271.462 & 6 July 1996 & F555W & 7400 \nl
2450616.593 & 16 June 1997 & F814W & 7400 \nl
2450616.761 & 16 June 1997 & F555W & 4800 \nl
2450647.036 & 17 July 1997 & F814W &7400 \nl
2450647.204 & 17 July 1997 & F555W & 4800 \nl
\enddata
\end{deluxetable}

\begin{deluxetable}{ccccccccc}
\tablecaption{Positions and Properties of Candidate Cepheids}
\tablehead{
\colhead{ID} & \colhead{x (pix)} & \colhead{y(pix)} & \colhead{$\bar{V}$} &
\colhead{$\bar{I}$} & \colhead{Amp.} & \colhead{$P$ (d)} &
\colhead{$V$ prob.} & \colhead{$I$ prob.}
}
\startdata
        37 & 476.14 &  68.22 & 27.45 & 26.03 & 0.68 & 30.00 & 0.75 & 0.12 \nl
        71 & 502.61 &  74.47 & 27.35 & 26.20 & 1.02 & 28.00 & 0.98 & 0.88 \nl
       200 & 139.98 &  98.63 & 27.46 & 26.55 & 0.89 & 54.00 & 0.81 & 0.32 \nl
       304 & 489.95 & 126.40 & 27.36 & 26.45 & 0.74 & 32.00 & 0.85 & 0.83 \nl
       307 & 520.29 & 126.66 & 27.32 & 26.53 & 0.62 & 33.00 & 0.69 & 0.68 \nl
       342 & 108.36 & 134.79 & 27.51 & 25.66 & 0.67 & 59.00 & 0.30 & 0.50 \nl
       391 & 132.02 & 146.45 & 27.01 & 26.42 & 0.73 & 28.00 & 0.55 & 0.91 \nl
       445 & 330.47 & 155.53 & 27.65 & 26.33 & 0.85 & 46.00 & 0.74 & 0.79 \nl
       500 & 186.48 & 169.34 & 27.96 & 26.69 & 0.75 & 30.00 & 0.88 & 0.87 \nl
       526 &  52.90 & 176.19 & 27.01 & 26.32 & 0.61 & 35.00 & 0.68 & 0.84 \nl
       722 & 576.78 & 230.69 & 27.24 & 27.20 & 0.62 & 39.00 & 0.78 & 0.01 \nl
       747 & 632.36 & 240.76 & 27.52 & 26.52 & 0.66 & 48.00 & 0.53 & 0.31 \nl
       774 & 406.08 & 245.75 & 27.43 & 26.46 & 0.73 & 24.00 & 0.53 & 0.33 \nl
       780 & 155.82 & 247.20 & 27.26 & 26.39 & 0.64 & 52.00 & 0.72 & 0.41 \nl
       871 & 543.66 & 264.40 & 27.29 & 25.97 & 0.73 & 48.00 & 0.88 & 0.84 \nl
       982 & 404.30 & 290.64 & 28.00 & 26.57 & 0.79 & 39.00 & 0.72 & 0.87 \nl
      1143 & 122.53 & 321.96 & 27.41 & 26.39 & 0.60 & 29.00 & 0.56 & 0.42 \nl
      1165 & 710.62 & 325.30 & 27.24 & 26.53 & 0.64 & 60.00 & 0.60 & 0.12 \nl
      1197 & 421.60 & 332.81 & 27.65 & 27.68 & 0.87 & 46.00 & 0.77 & 0.00 \nl
      1211 &  53.49 & 335.78 & 27.19 & 25.59 & 0.61 & 36.00 & 0.73 & 0.01 \nl
      1299 & 246.75 & 355.21 & 27.79 & 26.94 & 0.97 & 41.00 & 0.89 & 0.49 \nl
      1326 & 191.60 & 362.40 & 27.52 & 26.41 & 0.79 & 28.00 & 0.86 & 0.71 \nl
      1334 & 460.92 & 364.14 & 27.19 & 26.11 & 0.64 & 32.00 & 0.69 & 0.40 \nl
      1392 & 261.67 & 382.01 & 27.57 & 26.72 & 0.90 & 24.00 & 0.86 & 0.86 \nl
      1459 & 110.72 & 400.82 & 27.37 & 26.04 & 0.62 & 28.00 & 0.55 & 0.03 \nl
      1490 & 197.25 & 410.22 & 27.19 & 26.39 & 0.71 & 35.00 & 0.87 & 0.88 \nl
      1505 & 366.53 & 412.77 & 27.49 & 26.46 & 0.62 & 50.00 & 0.42 & 0.25 \nl
      1538 & 186.76 & 424.29 & 27.55 & 26.12 & 0.66 & 45.00 & 0.56 & 0.59 \nl
      1545 & 634.66 & 425.51 & 27.55 & 27.22 & 0.70 & 25.00 & 0.65 & 0.53 \nl
      1639 & 608.91 & 445.56 & 27.64 & 25.74 & 0.66 & 57.00 & 0.18 & 0.44 \nl
      1664 & 742.07 & 450.82 & 27.20 & 26.30 & 0.61 & 29.00 & 0.49 & 0.44 \nl
      1672 & 450.42 & 453.07 & 27.50 & 26.07 & 0.68 & 31.00 & 0.75 & 0.20 \nl
      1713 & 449.79 & 461.08 & 27.78 & 26.35 & 0.64 & 46.00 & 0.33 & 0.52 \nl
      1724 & 454.95 & 462.87 & 27.41 & 25.67 & 0.75 & 41.00 & 0.87 & 0.29 \nl
      1805 & 130.69 & 475.35 & 27.48 & 26.56 & 0.68 & 42.00 & 0.72 & 0.50 \nl
      1991 & 447.24 & 506.10 & 27.48 & 26.33 & 0.65 & 36.00 & 0.72 & 0.62 \nl
      2007 & 288.52 & 509.46 & 27.73 & 26.74 & 0.61 & 41.00 & 0.46 & 0.26 \nl
      2035 & 429.14 & 512.80 & 27.39 & 25.25 & 0.97 & 28.00 & 0.97 & 0.00 \nl
      2174 & 166.44 & 541.17 & 27.56 & 26.33 & 0.65 & 36.00 & 0.68 & 0.58 \nl
      2177 &  80.42 & 542.45 & 27.56 & 27.68 & 0.62 & 50.00 & 0.34 & 0.00 \nl
      2210 & 742.00 & 549.34 & 27.58 & 26.13 & 0.63 & 34.00 & 0.65 & 0.33 \nl
      2333 & 571.57 & 571.99 & 27.58 & 26.55 & 0.60 & 60.00 & 0.19 & 0.09 \nl
      2341 & 479.83 & 572.91 & 27.63 & 26.56 & 0.62 & 25.00 & 0.51 & 0.31 \nl
      2497 & 420.90 & 599.54 & 27.72 & 25.59 & 0.70 & 44.00 & 0.50 & 0.20 \nl
      2521 & 245.29 & 604.07 & 27.28 & 26.17 & 0.71 & 24.00 & 0.33 & 0.02 \nl
      2547 & 271.99 & 608.05 & 27.57 & 27.23 & 0.79 & 26.00 & 0.81 & 0.61 \nl
      2573 & 156.12 & 612.11 & 27.63 & 26.21 & 0.65 & 58.00 & 0.18 & 0.29 \nl
      2625 & 176.03 & 621.03 & 27.39 & 26.27 & 0.76 & 25.00 & 0.66 & 0.18 \nl
      2632 & 717.83 & 622.46 & 27.38 & 26.28 & 0.61 & 26.00 & 0.41 & 0.11 \nl
      2697 & 391.70 & 633.54 & 27.26 & 25.91 & 0.62 & 44.00 & 0.77 & 0.59 \nl
      2732 & 517.97 & 639.63 & 27.13 & 27.34 & 0.61 & 60.00 & 0.65 & 0.00 \nl
      2774 & 284.92 & 646.25 & 26.78 & 25.24 & 0.67 & 32.00 & 0.56 & 0.00 \nl
      2811 & 399.27 & 652.32 & 27.72 & 26.35 & 0.70 & 42.00 & 0.56 & 0.63 \nl
      2848 & 353.40 & 657.40 & 26.90 & 26.56 & 0.89 & 59.00 & 1.00 & 0.38 \nl
      2862 & 365.32 & 659.00 & 27.72 & 27.28 & 0.78 & 33.00 & 0.85 & 0.13 \nl
      2958 & 757.54 & 673.55 & 27.28 & 27.25 & 0.72 & 26.00 & 0.57 & 0.59 \nl
      2968 & 613.42 & 676.07 & 27.60 & 26.74 & 0.66 & 35.00 & 0.69 & 0.48 \nl
      2984 & 472.25 & 678.20 & 27.19 & 26.37 & 0.65 & 37.00 & 0.81 & 0.80 \nl
      3130 & 474.75 & 713.00 & 27.63 & 26.64 & 0.65 & 56.00 & 0.18 & 0.06 \nl
      3194 & 432.11 & 729.14 & 27.68 & 25.34 & 0.72 & 43.00 & 0.59 & 0.02 \nl
      3237 & 187.01 & 739.28 & 27.93 & 26.53 & 0.83 & 48.00 & 0.46 & 0.72 \nl
\enddata
\end{deluxetable}

\begin{deluxetable}{cccccc}
\tablecaption{Other Distance Estimates for NGC~4603 and Cen30}
\tablehead{
\colhead{Paper} & \colhead{Method} & \colhead{Number of} & \colhead{$cz_{Cen30,LG}$} & \colhead{$v_{4603}$} & \colhead{$v_{Cen30}$} \nl
\colhead{} & \colhead{} & \colhead{Galaxies} & \colhead{(\kms)} & \colhead{(\kms)} & \colhead{(\kms)} 
}
\startdata	
	Faber et al. 1989& $D_n-\sigma$ & 5& 2802 & \dots & $2221\pm208$ \nl
	Lucey and Carter 1988& $D_n-\sigma$ & 15& 2809& \dots & $3095\pm335 $ \nl
	Willick et al. 1997/EGAL& $D_n-\sigma$ & 22 & 2807 & \dots & $2917\pm155$ \nl
	Aaronson et al. 1989 & Forward TF & 10 & 2804 &2740 &$2830\pm248$ \nl
	Willick et al. 1997/HMCL& Forward TF & 10& 3139& 2759& $3445\pm213$ \nl
	Willick et al. 1997/HMCL& Inverse TF & 10& 3139& 2606& $3251\pm201$ \nl
	Willick et al. 1997/MAT& Forward TF & 5& 3228 & 2599& $3000\pm273$ \nl
	Willick et al. 1997/MAT& Inverse TF & 5& 3228 & 2443& $2881\pm262$ \nl
	Giovanelli et al. 1998& Forward TF & 39& 2783 & $2546 \pm 397$& $3012\pm98$ \nl
	This Work &Cepheid&1&2807&$2395 \pm 306$&$\ge 2395\pm 306$ \nl
\enddata
\end{deluxetable}
 
\newpage

\setcounter{figure}{9}

\begin{figure*}
\vskip9truein
\includegraphics{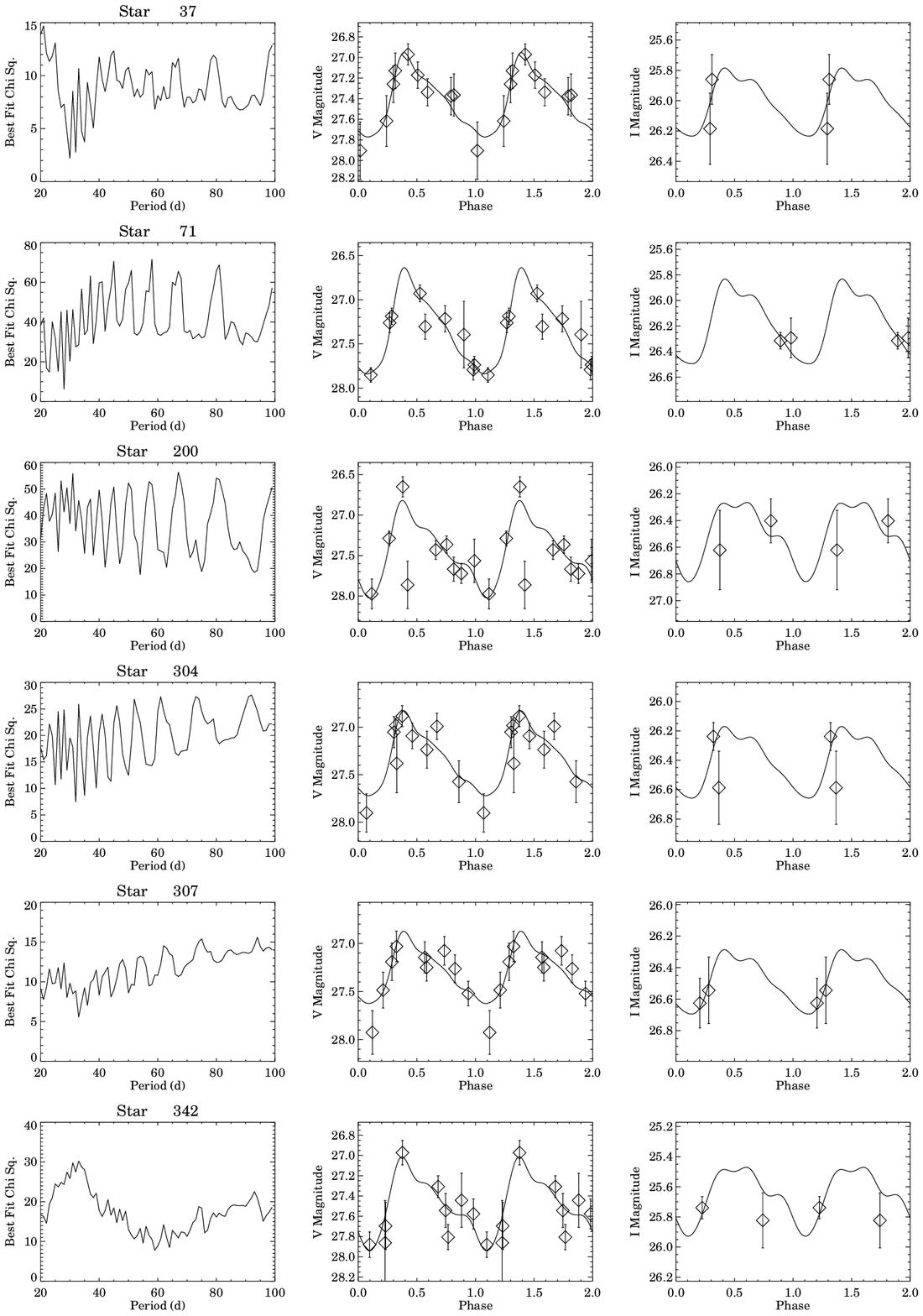} 
\caption[f10.ps]{Light curves of a subset of our candidate Cepheids.  In
the leftmost panel for each star, the variation of $\chi^2$ with period
for fits to template Cepheid light curves is plotted; aliasing is
readily apparent.  In the center, the $V$ magnitude for the candidate
is plotted as a function of phase over two cycles, along with the
best-fitting template light curve.  The rightmost panel shows a similar
plot for $I$ magnitudes.  Light curves for the entire sample may be obtained from our website.}
\end{figure*}

\end{document}